\documentclass[aps,prr,twocolumn,amsmath,amssymb,superscriptaddress,floatfix,superscriptaddress,longbibliography]{revtex4-2}

\usepackage{graphicx}
\usepackage{multirow}
\usepackage[dvipsnames]{xcolor}
\usepackage{bm}
\usepackage{amsfonts,amssymb,amsmath}
\usepackage{graphicx,dcolumn,bm,color,braket,slashed}
\usepackage{times} 
\usepackage{comment, ulem}

\def\ket#1{|#1\rangle }
\def\bra#1{\langle #1 |}

\begin{document}

\title{Higher-Order Topological Systems and Their Sub-Symmetry-Protected Topology}

\author{Myungjun Kang}
\affiliation{Department of Physics, Hanyang University, Seoul 04763, Korea}
\affiliation{High Pressure Research Center, Hanyang University, Seoul 04763, Korea}

\author{Wonjun Sung}
\affiliation{Department of Physics, Hanyang University, Seoul 04763, Korea}

\author{Sonu Verma}
\affiliation{Center for Theoretical Physics of Complex Systems, Institute for Basic Science (IBS) Daejeon 34126, Republic of Korea}
\affiliation{Department of Physics, Rheinland-Pf{\"a}lzische Technische Universit{\"a}t Kaiserslautern-Landau, Kaiserslautern 67663, Germany}

\author{Sangmo Cheon}
\email{sangmocheon@hanyang.ac.kr}
\affiliation{Department of Physics, Hanyang University, Seoul 04763, Korea}
\affiliation{Research Institute for Natural Science, Hanyang University, Seoul 04763, Korea}

\begin{abstract}
Symmetry and topology are essential principles in topological physics.
Recently, the idea of sub-symmetry-protected topology---where some of the original symmetries are broken while a remaining subset, called sub-symmetries, continues to protect specific boundary states---has been developed.
Here, we extend sub-symmetry-protected topology to higher-order topological systems from second-order topological insulators to semimetals.
By introducing a sub-symmetry-protecting perturbation that acts on a single sublattice and selectively preserves specific topological boundary states, we track the evolution of these states and their topological features using numerical and analytical methods, and we show that state-resolved quadrupole moments diagnose which corner or hinge modes remain topological.
As a representative example of a second-order topological insulator, we begin with the Benalcazar–Bernevig–Hughes model.
We demonstrate that, under a sub-symmetry-protecting perturbation, sub-symmetry-protected corner states remain pinned at zero energy and maintain  quantized state-resolved quadrupole moments.
In contrast, corner states on sub-symmetry-broken boundaries shift away from zero energy and lose their quantized character.
We further extend this framework to a three-dimensional second-order topological semimetal, constructed by stacking second-order topological insulator layers, and analyze how second-order Fermi arc states---hinge-localized modes that link the projections of bulk Dirac points, in contrast to conventional surface Fermi arcs---evolve under a sub-symmetry-protecting perturbation.
While one second-order Fermi arc becomes dispersive and loses its quadrupolar character under a sub-symmetry-breaking perturbation, the 
remaining second-order Fermi arcs retain chiral symmetry and preserve quantized quadrupolar characters.
These findings demonstrate that sub-symmetry-protected topology can manifest in both insulating and gapless phases, offering routes to engineering symmetry-resilient topological phases in electronic, photonic, and synthetic systems.
\end{abstract}

\maketitle

\newpage
\section{Introduction}
The topological phase of a system, along with its associated boundary states, is protected by topology-protecting symmetry and governed by the bulk-boundary correspondence~\cite{hasan2010colloquium}.
When the bulk possesses a nontrivial topological invariant, the boundary of a finite system necessarily accommodates robust boundary states that remain stable against any perturbation preserving the topology-protecting symmetry.
Prototypical examples of such nontrivial topological invariants include the quantized electric polarization (or Berry phase)~\cite{resta1994modern,rabe2007physics,qi2008topological}, as well as the $\mathbb{Z}_2$ topological index proposed by Fu and Kane~\cite{fu2007topological}.

Recently, there has been growing interest in higher-order topological systems, where topological boundary states appear at codimension greater than one—that is, at dimensions lower than those of conventional edge or surface states~\cite{schindler2018higher}.
An $n$th-order topological phase in $d$ dimensions hosts boundary modes of dimension $(d-n)$, and the associated topological index depends on the order.
For instance, the previous $\mathbb{Z}_2$ index for topological insulators indicates that the second-order topological insulating phase is trivial, and the $\mathbb{Z}_4$ index is required to see the bulk-boundary correspondence~\cite{khalaf2018symmetry}.

The concept of higher-order topology has been extended to gapless systems, giving rise to unconventional boundary phenomena beyond conventional surface states.
Such an extension is exemplified by first-order Dirac semimetals, a prototypical class of topological semimetals~\cite{young2012dirac,yang2014classification}.
These systems are characterized by Dirac cones in the bulk, while the surface projections of the Dirac points are connected via surface-localized Fermi arc states.
In second-order topological semimetals (SOTSMs), the surface-localized Fermi arcs of conventional Dirac semimetals evolve into second-order Fermi arcs, which are no longer found on the surface but instead appear along the hinges in three-dimensional systems~\cite{wieder2020strong}. 

\begin{figure}[t]
\includegraphics[width=0.5\textwidth]{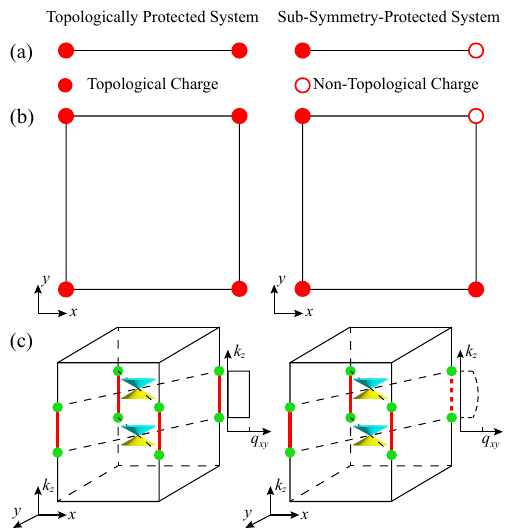}
\caption{\label{fig0:Schematics}
\textbf{Schematics of the topological systems and their boundary states under sub-symmetry-protecting perturbation.}
Left (right) panels show the cases without (with) sub-symmetry-protecting perturbations.
\textbf{(a)} One-dimensional topological insulator.
Edge states with (without) topological charge are shown as filled (open) red circles.
\textbf{(b)} Two-dimensional SOTI with corner states.
Corner states protected (unprotected) by sub-symmetry exhibit quantized (non-quantized) quadrupole moments, shown by the filled (open) red circles.
\textbf{(c)} Three-dimensional SOTSM and the corresponding second-order Fermi arc states.
Green circles represent the surface projections of bulk Dirac points, connected by second-order Fermi arc states (red lines). The topological character of each second-order Fermi arc is distinguished by the quadrupole moment along $k_z$, as shown in the inset. The second-order Fermi arc with a non-quantized quadrupole moment is represented by a dotted red line, with its non-quantized quadrupole value also indicated in the inset.
}
\end{figure}

Figure~\ref{fig0:Schematics}(c) schematically illustrates a three-dimensional SOTSM, where red second-order Fermi arc states connect the green Dirac point projections along the hinges.
Several proposals have demonstrated that such second-order Fermi arc states can be realized by stacking two-dimensional second-order topological insulators (SOTIs), resulting in SOTSM phases~\cite{wieder2020strong,wang2021engineering,chen2023quasicrystalline}.

In realistic systems, imperfections such as disorder, boundary roughness, and fabrication asymmetries often break the ideal symmetries assumed to protect topological phases.
This raises the question of whether boundary-localized topological states can survive under partial symmetry breaking.
Rather than simply tolerating such imperfections, recent advances in material design and synthetic platforms offer the possibility of selectively engineering symmetries to stabilize desired boundary states.
This perspective reframes symmetry breaking as a design principle rather than a constraint, and motivates the study of sub-symmetry-protected topology~\cite{wang2023sub,kang2024subsymmetry,verma2024non}.
Sub-symmetry-protected topology reveals that even when global symmetries are broken, residual sub-symmetries may suffice to preserve topological boundary modes---enabling new routes for symmetry-controlled localization and manipulation.

Here, we extend the concept of sub-symmetry-protected topology to higher-order topological systems, introducing perturbations that selectively preserve only a subset of the original symmetry---termed sub-symmetry-protecting perturbations.
The central idea of sub-symmetry-protected topology is to treat the full topology-protecting symmetry as a composite of sub-symmetries, each of which may independently protect specific boundary states even when the full symmetry is broken.
When the global topology-protecting symmetry is partially broken---preserving only a subset of its constituent sub-symmetries---certain boundary states can remain topologically protected. These are the so-called sub-symmetry-protected boundary states, stabilized by the residual symmetries.
Crucially, such selective protection enables spatial localization of topological modes on specific boundaries of a finite system.

Experimental realizations of sub-symmetry-protected boundary states have been demonstrated in optical implementations of one-dimensional topological insulators and two-dimensional SOTIs~\cite{wang2023sub}. Their topological nature has been corroborated through quantized polarization measurements in one-dimensional systems~\cite{kang2024subsymmetry}, and theoretically supported by non-Bloch band theory~\cite{verma2024non}. The sub-symmetry-protected framework has also been extended to superconducting systems, giving rise to sub-symmetry-protected Majorana modes~\cite{kang2024subsymmetry}.
Despite these developments, several important questions remain open. In particular, a topological index characterizing sub-symmetry-protected boundary states in higher-order topological systems has not yet been established. 
Furthermore, the extension of sub-symmetry-protected concepts to gapless systems such as topological semimetals remains largely unexplored. These gaps highlight a rich and largely unexplored realm of symmetry-resilient boundary phenomena that this work aims to address.

In this work, we demonstrate the realization of sub-symmetry-protected topology in a two-dimensional SOTI and a three-dimensional SOTSM.
We analyze the evolution of their boundary states and associated topological features using complementary numerical calculations and an analytical Jackiw–Rebbi theory.
We begin by analyzing the Benalcazar–Bernevig–Hughes (BBH) model as a representative SOTI system, where the topology-protecting symmetry is the chiral symmetry~\cite{benalcazar2017quantized}.
Under a sub-symmetry-protecting perturbation, this model exhibits zero-energy corner states localized at three of the four corners of a finite geometry.
Since the topology-protecting symmetry is broken, the conventional bulk-boundary correspondence fails to capture the topological nature of the corner states. 
In earlier studies of sub-symmetry-protected topology, one-dimensional sub-symmetry-protected boundary states were characterized by
topological charges---such as quantized dipole moments or end-charge distributions of the boundary states---as illustrated in Fig.~\ref{fig0:Schematics}(a)~\cite{wang2023sub,kang2024subsymmetry}.
Generalizing this idea to higher-order systems, we define the topological invariant of a higher-order boundary state as a state-resolved quadrupole character because the sub-symmetry-protected boundary states are localized at a specific sublattice site due to the chiral symmetry.
A state-by-state analysis reveals that for the BBH model, certain corner states retain quantized quadrupole moments---and hence their topological character---even without the full symmetry.
In contrast, the remaining corner state, located at a boundary where the relevant sub-symmetry is broken, shifts away from zero energy and loses its quantized topological value~[Fig.~\ref{fig0:Schematics}(b)].

Expanding to semimetallic systems, we construct a three-dimensional SOTSM by stacking layers of the BBH model~\cite{wang2021engineering}, each subjected to an identical sub-symmetry-protecting perturbation.
This layered configuration yields second-order Fermi arc states with distinctive behavior: while the second-order Fermi arc spectrum localized on the sub-symmetry-broken hinge becomes curved and perturbed, the spectra on the sub-symmetry-protected hinges remain flat and robust. As a result, the perturbed hinge exhibits a non-quantized quadrupole moment along the momentum axis connecting the two Dirac points, whereas the sub-symmetry-protected hinges maintain quantized quadrupole momenta~[Fig.~\ref{fig0:Schematics}(c)].
The bulk Dirac points remain unaffected by the sub-symmetry-protecting perturbation due to the stacking characteristics of the system.
Our findings demonstrate that sub-symmetry protection can be successfully extended to higher-order topological systems, and they pave the way for realizing sub-symmetry-protected phenomena in a wide range of physical platforms, including tunable electronic materials~\cite{drost2017topological,huda2020tuneable} and engineered photonic systems~\cite{meier2016observation,ozawa2019topological}.

This paper is organized as follows: In Sec. II, we introduce the BBH model as a representative SOTI and demonstrate the emergence of sub-symmetry-protected corner states under sub-symmetry-protecting perturbations.
We analyze the corresponding topological features using numerical spectra, Jackiw–Rebbi boundary solutions, and state-resolved quadrupole-moment diagnostics.
In Sec. III, we extend our analysis to a three-dimensional SOTSM constructed by stacking BBH layers, and investigate the evolution of second-order Fermi arc states under sub-symmetry-protecting perturbations.
In the Conclusion, we summarize our findings and discuss potential implications of sub-symmetry-protected topology in tunable and engineered physical systems.
Appendix~\ref{Appendix:Surface} presents a low-energy boundary theory that describes the emergence of sub-symmetry-protected corner modes in SOTIs, and Appendix~\ref{Appendix:Edge} provides an effective edge theory for interpreting sub-symmetry-protected protection in one-dimensional subsystems.

\begin{figure*}[t]
\includegraphics[width=\textwidth]{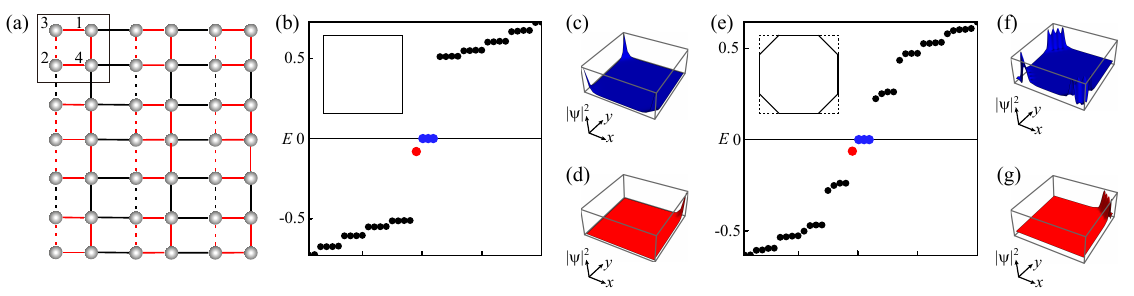}
\caption{\label{fig1:2DQI}
\textbf{Sub-symmetry-protected corner states in the BBH model.}
\textbf{(a)} Schematics of the BBH model.
Each square unit cell contains four sublattice sites labeled 1 through 4.
Black (intercell, $\lambda$) and red (intracell, $\gamma$) lines indicate hopping amplitudes; dotted lines denote negative amplitude.
\textbf{(b)} Energy spectrum of finite square geometry.
Three zero-energy sub-symmetry-protected states appear (blue), along with one finite-energy trivial state (red).
The inset illustrates the system's geometry, which is a square.
The parameters are $\gamma=-0.5, \lambda=1$, and $m_1=0.2$.
The system is finite in the $xy$ plane ($20 \times 20$ unit cells).
The total number of states is $2n=1600$.
\textbf{(c,d)} Spatial distributions of the wavefunctions corresponding to the sub-symmetry-protected (c, blue) and the trivial (d, red) corner states.
\textbf{(e)} Energy spectrum for a modified system with corner-cut geometry. 
The inset shows the geometry, with dashed lines indicating the regions that have been removed. 
Each removed region is comprised of a total of six unit cells.
The sub-symmetry-protected states (blue) remain at zero energy, while the trivial state (red) is lifted.
The parameters are identical to panel (b), and the total number of states is $2n=1504$.
\textbf{(f,g)} Wavefunction distributions of the sub-symmetry-protected (f, blue) and the trivial (g, red) in-gap states in the corner-cut geometry.
The sub-symmetry-protected states retain their localization, demonstrating robustness against geometric deformation and confirming their geometry-independent topological origin.
}
\end{figure*}

\section{Second-Order Topological Insulator}
\subsection{Model and Symmetry}
We begin by examining how a sub-symmetry-protecting perturbation affects a two-dimensional SOTI. As a representative example, we consider the Benalcazar–Bernevig–Hughes (BBH) model~\cite{benalcazar2017quantized}, a spinless quadrupole insulator known to host four topologically protected corner states, $\psi_i$ ($i=1,2,3,4$).
The schematics can be seen in Fig.~\ref{fig1:2DQI}(a), and the tight-binding Hamiltonian takes the form
\begin{eqnarray}
\label{eq:tb}
    H_{\text{BBH}} &=&\sum_{i,j} \gamma c^\dagger_{1,i,j} \left(c_{4,i,j}+c_{3,i,j}\right)\\
&&~~~+\gamma c^\dagger_{2,i,j} \left(c_{4,i,j}-c_{3,i,j}\right)\nonumber\\
&&~~~+\lambda c^\dagger_{1,i,j}\left(c_{4,i,j+1}+c_{3,i+1,j}\right)\nonumber\\
&&~~~+\lambda c^\dagger_{2,i,j}\left(c_{4,i-1,j}-c_{3,i,j-1}\right)+h.c.,\nonumber
\end{eqnarray}
where $c_{n,i,j}^\dagger$/$c_{n,i,j}$ denotes the creation/annihilation operator acting on the $n$th sublattice site in the unit cell located at $(i,j)$ position in the $xy$ plane.
The parameters $\gamma$ and $\lambda$ indicate the intracell and intercell hopping terms, respectively.
The system is in a topological phase when $|\gamma|<|\lambda|$ and the topology-protecting symmetry is the chiral symmetry~\cite{benalcazar2017quantized}.

We first examine the topology-protecting symmetry and sub-symmetries of the BBH model~\cite{benalcazar2017quantized,wang2023sub}.
Considering the Hamiltonian of the BBH model $H_\text{BBH}$ given in Eq.~(\ref{eq:tb}), it possesses a chiral symmetry,
\begin{eqnarray}
    \left\{\Gamma, H_{\text{BBH}}\right\}  = 0,
\end{eqnarray}
for $\Gamma^\dagger=\Gamma,\ \Gamma^2=\mathbf{1}$.
This means that $\Gamma$ flips the energy of any eigenstate. Concretely, if $H_{\text{BBH}}\psi_E=E\psi_E$, then
\begin{eqnarray}\label{eq:pairing}
H_{\text{BBH}}(\Gamma\psi_E)=-\,\Gamma H_{\text{BBH}}\psi_E=-E(\Gamma\psi_E),
\end{eqnarray}
so the spectrum is symmetric about zero and energy levels occur in $\pm E$ pairs.
Moreover, zero-energy states have special properties. They are mapped to themselves (up to a phase) by $\Gamma$ and can therefore be taken as simultaneous eigenstates $\Gamma\psi_0=\gamma\psi_0$ with $\gamma=\pm1$.

In the BBH basis, the chiral operator is diagonal across the sublattice sectors and can be written as
\begin{eqnarray}
\Gamma=P_1+P_2-P_3-P_4 ,
\end{eqnarray}
where $P_i$ projects onto sublattice $i$.
This shows that the $\Gamma = \pm 1$ sectors are spanned by sublattice sectors---i.e., the collections of sublattice sites---$\{1,2\}$ and $\{3, 4\}$, respectively. 
Thus, a zero-energy mode can be chosen to live entirely within one of these sector.
In the BBH geometry, the four corner zero modes are localized on definite sublattices determined by the boundary termination. 
The intra-/inter-cell dimerization sets the sign of the boundary mass in the Jackiw–Rebbi picture, so each corner aligns with either the $\Gamma=+1$ sector $\{1,2\}$ or the $\Gamma=-1$ sector $\{3,4\}$, as will be discussed later.

As an example, we consider a sub-symmetry-protecting perturbation which only acts on sublattice $1$,
\begin{eqnarray}
    V_{\text{SSP}}=\sum_{i,i',j,j'} v_{i,i',j,j'} ~ c^\dagger_{1,i,j}c_{1,i',j'},
\end{eqnarray}
where $v_{i,i',j,j'}$ is a perturbation parameter.
Then, we consider the total Hamiltonian as $H=H_{\text{BBH}}+V_{\text{SSP}}$. 
The global chiral anticommutation is broken since
\begin{eqnarray}\label{eq:broken-global}
    \Gamma H\Gamma^{-1}=\Gamma(H_{\text{BBH}}+V_{\text{SSP}})\Gamma^{-1} =-H+2V_{\text{SSP}} .
\end{eqnarray}
However, on the complementary subspace 
\begin{eqnarray}
    Q=\mathbf{1}-P_1 ,
\end{eqnarray}
spanned by sublattice sites $i=2,3,4$, the perturbation vanishes, $V_{\text{SSP}}Q=0$, and Eq.~\eqref{eq:broken-global} reduces to the projected chiral anticommutation relation
\begin{eqnarray}\label{eq:partial-chiral}
    \Gamma H\Gamma^{-1}Q=-HQ.
\end{eqnarray}
This also can be written as 
\begin{eqnarray}
    \{\Gamma_Q,\,H_Q\}=0,
\end{eqnarray}
where $\Gamma_Q\equiv Q\Gamma Q$ and $H_Q\equiv QHQ$.
Hence, for any eigenstate $\phi$ supported in $Q$ subspace, the $\pm E$ pairing Eq.~\eqref{eq:pairing} still holds with $H$ in place of $H_{\text{BBH}}$, implying that such boundary modes remain pinned at zero energy when protected by the sub-symmetry.
By contrast, states with support on $P_1$ do not obey Eq.~\eqref{eq:partial-chiral} and are generically shifted away from zero energy.

This is precisely the sense in which we use the term sub-symmetry, an exact symmetry that acts faithfully on a proper subspace, sufficient to protect a subset of boundary modes even when the global symmetry is broken~\cite{wang2023sub}.
In the following detailed numerical and analytical diagnostics for the BBH model, we will show that three corner modes are supported on sublattices $i=2,3,4$ and thus remain at $E=0$ under the sub-symmetry-protecting perturbation, while the corner mode tied to sublattice~1 is unprotected and moves away from zero energy.

\subsection{Numerical and Analytical Analysis}
To explicitly implement a sub-symmetry-protecting perturbation that partially breaks the topology-protecting symmetry, we introduce the following quasi-periodic on-site potential~\cite{roy2021reentrant,roy2023critical}:
\begin{eqnarray}
\label{eq:SOTIpert}
    m(i,j) = m_1 \cos\left[\pi\left(\sqrt{5}-1\right)(i+j)\right].
\end{eqnarray}
This potential is applied exclusively to sublattice-1 within each unit cell.
Here, $i$ and $j$ indicate the position of the unit cell along the $x$ and $y$ directions, respectively, and $m_1$ sets the perturbation strength.

Figure~\ref{fig1:2DQI}(b) displays the numerically computed energy spectrum of a finite system with square geometry, as shown in the inset. The spectrum exhibits three in-gap states at zero energy (blue), corresponding to sub-symmetry-protected corner states, and one additional in-gap state at non-zero energy (red), identified as topologically trivial.
Figures~\ref{fig1:2DQI}(c) and (d) display the spatial distribution of the wavefunctions associated with the sub-symmetry-protected and trivial in-gap states, respectively.
Each of the three sub-symmetry-protected states exhibits strong localization at a distinct corner, predominantly on the $2$, $3$, or $4$ sublattice site of each unit cell, respectively.
Due to such localization at a specific sublattice site, these sub-symmetry-protected states remain as eigenstates of the chiral operator and are protected by the sub-symmetry.
In contrast, the trivial in-gap state---originating from the perturbed $1$ sublattice---shows a delocalized profile that extends across multiple sublattice sites, indicating the breakdown of topological protection at that corner.
Thus, this delocalized state is no longer an eigenstate of the chiral operator.
Although the spatial distribution in Fig.~\ref{fig1:2DQI}(d) may appear weak, numerical analysis confirms that the trivial state is delocalized over multiple sublattice sites, without sharp localization at any corner. This behavior is consistent with the Jackiw–Rebbi analytical solution, which predicts a non-topological edge state with finite weight on both sublattices when sub-symmetry protection is absent, as will be discussed below.

To analytically examine the effect of sub-symmetry-protecting perturbations on the corner states, we construct a low-energy effective boundary theory. This approach is based on an effective boundary Hamiltonian that captures the spatial structure and localization behavior of the boundary modes. In particular, Jackiw–Rebbi-type solutions of the effective boundary theory naturally describe the emergence and robustness of corner-localized zero modes~\cite{schindler2020dirac}. These solutions explicitly show how the perturbation modifies the boundary mass terms, thereby selectively preserving or destabilizing corner states depending on the local sub-symmetry structure.

We begin with the Bloch Hamiltonian of the bulk BBH model, which is given as 
\begin{eqnarray}
\label{eq:QI}
    H_{\text{BBH}}(\textbf{k}) &=& \left( \gamma+\lambda \cos k_x \right) \tau_x-\lambda \sin k_x \tau_y \sigma_z \\
&&-\left( \gamma+\lambda \cos k_y\right)\tau_y \sigma_y-\lambda \sin k_y \tau_y\sigma_x.\nonumber
\end{eqnarray}
Here, $\tau_i$ and $\sigma_j$ are Pauli matrices acting on the sublattice degrees of freedom, where the basis is ordered as $\psi^T = (c_1, c_2, c_3, c_4)$ corresponding to the four sublattice sites labeled 1 through 4 within each unit cell.
The Bloch Hamiltonian has time-reversal ($T=K$), particle-hole ($C=\tau_z K$), and chiral ($\Gamma=\tau_z$) symmetries, placing the system in the BDI class~\cite{schnyder2008classification,chiu2016classification}.
This Bloch Hamiltonian contains momentum terms involving $k_x$ and $k_y$, which become differential operators when the system is finite along the $x$ and $y$ directions, respectively, enabling a continuum description of spatially localized corner states.

To derive Jackiw-Rebbi solutions from an appropriate Dirac Hamiltonian, we first express the momentum components in the polar coordinates as $(k_x,k_y)=\left(k_\perp \cos\phi-k_\parallel \sin\phi,k_\perp \sin\phi+k_\parallel \cos\phi\right)$, where $\phi$ is the angle relative to the $x$ axis. Here, $k_\perp$ and $k_\parallel$ are the perpendicular and tangential momentum components, respectively.
Taking linear approximation with respect to the momentum near the $\Gamma$ point for the Bloch Hamiltonian in Eq.~(\ref{eq:QI}) and setting $\alpha \equiv \gamma+\lambda$ for simplicity, the resulting low-energy effective Hamiltonian reads
\begin{eqnarray}
\label{eq:QID}
    H_{\text{BBH}}^\text{eff}(k_\perp, k_\parallel)  &=& -\lambda k_\perp \tau_y\left( \cos\phi \sigma_z+\sin\phi\sigma_x\right)\\
&&+\lambda k_\parallel \tau_y\left(\sin\phi \sigma_z-\cos\phi\sigma_x\right)\nonumber\\
&&+\alpha \left(\tau_x-\tau_y \sigma_y\right)\nonumber.
\end{eqnarray}

We consider a circular boundary in real space, where the momentum perpendicular to the edge is replaced by a differential operator, $k_\perp=-i\partial_r$, along the radial direction, while $k_\parallel$ remains a good quantum number~\cite{schindler2020dirac}. For simplicity, we set $\lambda=1$ without loss of generality. The corresponding effective edge Hamiltonian is given by
\begin{eqnarray}
    H_{\text{BBH}}^{\text{edge}} = k_\parallel \sigma_y + \alpha(\phi)\sigma_x,
    \label{eq:H_QI_surface}
\end{eqnarray}
where $k_\parallel=-i\partial_\phi$ is the differentical operator along edge,  $\sigma_i$ are Pauli matrices acting in the edge basis, and $\alpha(\phi)$ represents the angular-dependent Dirac mass. Explicitly,
\begin{eqnarray}
    \alpha(\phi) = \pm \alpha \left(\cos\phi \mp \sin\phi \right),
\end{eqnarray}
with the choice of signs depending on the angular sector. 
The relative sign of the mass term alternates as a function of the angular sector: $(+,-)$ for $0 \leq \phi \leq \pi/2$, $(-,+)$ for $\pi/2 \leq \phi \leq \pi$, $(-,-)$ for $\pi \leq \phi \leq 3\pi/2$, and $(+,+)$ for $3\pi/2 \leq \phi \leq 2\pi$.
Analyzing the spectrum of $H_{\text{BBH}}^{\text{edge}}$, we find that zero-energy solutions emerge precisely at angular positions $\phi=\pi/4,\,3\pi/4,\,5\pi/4,\,7\pi/4$, where the Dirac mass term in Eq.~(\ref{eq:H_QI_surface}) vanishes and changes sign. These points correspond to domain walls along the circular boundary, signaling the presence of corner-localized modes.  

In the presence of the sub-symmetry-protecting perturbation, however, the edge Hamiltonian acquires an additional mass contribution. This term vanishes at all corners except at $\phi=\pi/4$, where it remains finite~[Fig.~\ref{fig4:DiracMass}(b)]. As a consequence, the corner mode at $\phi=\pi/4$ is lifted from zero energy, while the other three corners at $\phi=3\pi/4,\,5\pi/4,\,7\pi/4$ retain robust zero-energy states. The detailed derivation is presented in Appendix~\ref{Appendix:Surface}.

To analyze the corner states, we reconsider the low-energy effective Hamiltonian of Eq.~(\ref{eq:QID}) in a real-space square geometry.
In the same polar coordinate system, we note that the tangential momentum $k_\parallel$ is no longer a good quantum number and vanishes at the corners.
Inserting the sub-symmetry-protecting perturbation and setting $k_\perp=-i\partial_r,k_\parallel=0$ and $\lambda=1$, the Dirac-type effective Hamiltonian of Eq.~(\ref{eq:QID}) becomes
\begin{eqnarray}
\label{eq:DiracBBH}
    H_{\text{BBH}}^\text{Dirac} &=& i \partial_r \tau_y\left( \cos\phi \sigma_z+\sin\phi\sigma_x\right)\\
&&+\alpha \left(\tau_x-\tau_y \sigma_y\right)+m(r) \left(\tau_0+\tau_z\right)\left(\sigma_0+\sigma_z\right)\nonumber.
\end{eqnarray}
Here, $m(r)$ is the sub-symmetry-protecting perturbation term, which is the approximated quasi-periodic on-site potential in the continuum limit.
This perturbation term breaks the particle-hole and chiral symmetries but preserves the time-reversal symmetry.
The resulting Dirac-type effective Hamiltonian admits zero-energy Jackiw–Rebbi solutions localized at the corners~\cite{jackiw1976solitons,jackiw1981solitons}, regardless of the detailed form of $m(r)$. The solutions are given by
\begin{eqnarray}
    \psi_{\phi=\frac{5\pi}{4}}(r)=&\mathcal{N}e^{\sqrt{2}\alpha (r-R)}
	\begin{pmatrix}
	0\\
	1\\
	0\\
	0\\
	\end{pmatrix},\\
    \psi_{\phi=\frac{3\pi}{4}}(r)=&\mathcal{N}e^{\sqrt{2}\alpha (r-R)}
	\begin{pmatrix}
	0\\
	0\\
	1\\
	0\\
	\end{pmatrix},\\
    \psi_{\phi=\frac{7\pi}{4}}(r)=&\mathcal{N}e^{\sqrt{2}\alpha (r-R)}
	\begin{pmatrix}
	0\\
	0\\
	0\\
	1\\
	\end{pmatrix},
\end{eqnarray}
where $R$ is the distance from the center to the corner, $0\leq r \leq R$, and $\mathcal{N}$ is a normalization constant, respectively.
These analytical corner modes are in excellent agreement with the sub-symmetry-protected corner states obtained from the tight-binding calculation shown in Fig.~\ref{fig1:2DQI}(c) and symmetry analysis in the previous subsection.

To analytically capture the behavior of the trivial in-gap state with non-zero energy at $\phi=\frac{\pi}{4}$, we simplify the perturbation by setting a constant as $m(r)=m_1$.
In this case, we find the analytical solution for the corner state at $\phi=\frac{\pi}{4}$ as
\begin{eqnarray}
    \psi(r)=\mathcal{N}e^{\beta (r-R)}
	\begin{pmatrix}
	2\alpha+\sqrt{2}\beta\\
	0\\
	E-m_1\\
	E-m_1\\
	\end{pmatrix},
\end{eqnarray}
where $\beta=\sqrt{2\alpha^2+E(m_1-E)}$ and $\mathcal{N}$ is the normalization constant, respectively.
This solution agrees well with the numerical wavefunction profile shown in Fig.~\ref{fig1:2DQI}(d), confirming the topologically trivial nature of the perturbed corner state.

An alternative perspective is to analyze each edge of the system individually, rather than adopting a global polar-coordinate approach. In this framework, the edge of a SOTI can be modeled as a one-dimensional topological insulator~\cite{benalcazar2017quantized,schindler2020dirac}.
For the BBH model, the edge Hamiltonians for the four boundaries can be explicitly constructed in Cartesian coordinates. Each of these one-dimensional systems falls into the BDI symmetry class, characterized by the presence of chiral symmetry.
In the topological regime of $|\gamma|<|\lambda|$, each edge Hamiltonian is characterized by a non-zero chiral winding number, indicating that the edges function as one-dimensional topological insulators. However, the introduction of a sub-symmetry-protecting perturbation explicitly breaks chiral symmetry along the top and right edges, leading to the absence of a well-defined topological index in those directions.
Nevertheless, the resulting one-dimensional systems along the top and right edges continue to host sub-symmetry-protected boundary states—zero-dimensional corner modes—analogous to those found in the sub-symmetry-protected Su–Schrieffer–Heeger chain~\cite{kang2024subsymmetry}.
These “edges of edges” features are schematically illustrated in Fig.\ref{fig0:Schematics}(b), and their topological origin is further clarified through analytical calculations in Appendix~\ref{Appendix:Edge}.
The results are summarized in Table~\ref{table1}.
While each boundary can be formally mapped onto a one-dimensional topological insulator characterized by a chiral winding number, the emergence of corner-localized states and the associated topological features fundamentally derive from the second-order topology of the two-dimensional bulk.

\begin{table}[t]
\begin{tabular}{c c c c c c c} 
 \hline
 ~~Edge~~& ~~~~~~~~~$T$~~~~~~~~~ & ~~~~~~~~~$C$~~~~~~~~~ & ~~~~~~~~~$\Gamma$~~~~~~~~~ &~~~~~~~~~Class~~~~~~~~~\\
 \hline
top & 1 & 0 & 0 & AI\\ 
bottom & 1 & 1 & 1 & BDI\\
left & 1 & 1 & 1 & BDI\\
right & 1 & 0 & 0 & AI\\
 \hline
\end{tabular}
\caption{
\textbf{Symmetry content and Altland-Zirnbauer classes of effective edge Hamiltonians under a sub-symmetry-protecting perturbation.}
Entries indicate presence (1) or absence (0) of time-reversal $T$, particle--hole $C$, and chiral $\Gamma$ symmetries; for $T$ and $C$, $1$ denotes $T^{2}=+1$, $C^{2}=+1$.
The resulting Altland-Zirnbauer class for each edge is listed in the last column.
}
\label{table1}
\end{table}

To assess the geometry dependence of sub-symmetry and the resulting sub-symmetry-protected states, we examine the BBH model in a corner-cut geometry. Specifically, a right isosceles triangle composed of unit cells is removed from each of the four corners, as illustrated in the inset of Fig.~\ref{fig1:2DQI}(e).
Numerical calculations for this modified geometry reveal four in-gap states, each localized near a truncated corner. In the absence of perturbation, these states lie at zero energy, protected by the full topology-protecting symmetry of the system.
Upon introducing the sub-symmetry-protecting perturbation, three zero-energy in-gap states (blue) persist as sub-symmetry-protected corner states, while one additional in-gap state (red) shifts to finite energy and is topologically trivial, as shown in the energy spectrum in Fig.~\ref{fig1:2DQI}(e). This result demonstrates that the sub-symmetry-protected boundary states retain their topological character regardless of the specific boundary geometry.
As in the square-shaped system, the spatial distributions of the sub-symmetry-protected and trivial states in the corner-cut geometry are shown in Fig.~\ref{fig1:2DQI}(f,g), respectively. The sub-symmetry-protected states remain sharply localized, whereas the trivial state is more delocalized.
This delocalization, although not visually prominent in Fig.~\ref{fig1:2DQI}(g), is confirmed numerically through its extended spatial profile across multiple sublattice sites.
These observations are in excellent agreement with the predictions from the boundary theory and the Jackiw–Rebbi solutions, confirming that the existence of sub-symmetry-protected states is governed by the underlying sub-symmetry rather than the precise shape of the system boundary.
This highlights the potential of sub-symmetry engineering as a powerful approach to realizing robust, geometry-independent topological modes.

\begin{figure}[t]
\includegraphics[width=0.5\textwidth]{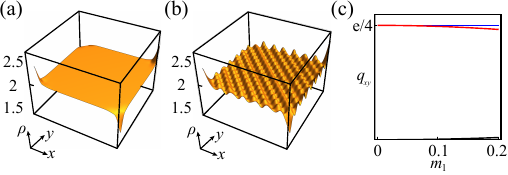}
\caption{\label{fig3:quadrupole}
\textbf{Charge density distribution of the occupied states and quadrupole moment with respect to the sub-symmetry-protecting perturbation.}
\textbf{(a,b)} Charge density distribution ($\rho$) for the occupied states (a) without and (b) with the sub-symmetry-protecting perturbation ($m_1=0.2$).
In (a), the bulk has a uniform charge, and the corner states have an identical absolute charge distribution.
For (b), the bulk charge distribution is not uniform and will result in a non-zero bulk quadrupole moment.
\textbf{(c)} State-resolved quadrupole moments of the bulk and corner states as a function of $m_1$.
The black line indicates the bulk quadrupole moment $q_{xy}^\text{bulk}$.
The red and blue lines correspond to the absolute value of $q_{xy}^i$ for the trivial and sub-symmetry-protected corner states, respectively.
The parameters excluding $m_1$ and geometry are the same as Fig.~\ref{fig1:2DQI}(b).
}
\end{figure}

\subsection{Topological Invariants}
The modern theory of polarization is used to examine the topological properties of the corner states.
For the BBH model, the topological nature can be captured by the quantization of the quadrupole moment of the system~\cite{resta1994modern,rabe2007physics}.
We define the state-resolved quadrupole moment $q_{xy}^i$ for the $i$th corner state in regards to electron charge $e$ as
\begin{eqnarray}
\label{eq:quad}
    q_{xy}^i
    &=&\frac{e}{4}\sum_{j,k}\left|\psi^i_{1, j,k}\right|^2+\left|\psi^i_{2,j,k}\right|^2\\
    && ~~~~~~~ -\left|\psi^i_{3,j,k}\right|^2-\left|\psi^i_{4,j,k}\right|^2,\nonumber
\end{eqnarray}
where $\psi^i_{n, j, k}$ denotes the amplitude of a wavefunction of the $i$th in-gap state at the $n$th sublattice site in the unit cell located at the $(j,k)$ position.
The same can be applied to the bulk states, where the bulk contribution to the quadrupole moment is the sum of all valance band states.
The bulk quadrupole moment is given as
\begin{eqnarray}
\label{eq:quad_blk}
    q_{xy}^\text{bulk}
    &=&\frac{e}{4}\sum_{m}\sum_{j,k}\left|\psi^m_{1,j,k}\right|^2+\left|\psi^m_{2,j,k}\right|^2\\
    && ~~~~~~~~~~~~~~-\left|\psi^m_{3,j,k}\right|^2-\left|\psi^m_{4,j,k}\right|^2,\nonumber
\end{eqnarray}
where $\psi^m_{n, j, k}$ denotes the amplitude of the $m$th occupied valance wavefunction the $n$th sublattice site in the unit cell located at the $(j,k)$ position.
The total quadrupole moment of the system is therefore given as
\begin{eqnarray}
\label{eq:quad_tot}    q_{xy}^\text{tot}=q_{xy}^\text{bulk}+\sum_{i}q_{xy}^i.
\end{eqnarray}
In the absence of sub-symmetry-protecting perturbation, the quadrupole moment of the system is quantized at $q_{xy}^\text{tot}=\frac{e}{2}$.
This comes from a zero quadrupole moment from the bulk, and two corner states~\cite{benalcazar2017quantized}.
This can be seen via the charge distribution of Fig.~\ref{fig3:quadrupole}(a), where there is a uniform distribution for the bulk and localized corner charges.

Figure~\ref{fig3:quadrupole}(b,c) summarizes the response to a sub-symmetry-protecting perturbation of strength $m_{1}$.
The sub-symmetry-protecting perturbation term conserves total charge of the system but redistributes it through the bulk and edge~[Fig.~\ref{fig3:quadrupole}(b)], thereby generating a finite bulk quadrupole $q^{\mathrm{bulk}}_{xy}(m_{1})$ that grows with $m_{1}$~[black line in Fig.~\ref{fig3:quadrupole}(c)].
Concomitantly, the corner mode on the sublattice not protected by the sub-symmetry loses its topological character.
Its state-resolved quadrupole $q^{i}_{xy}$ drifts continuously away from the quantized value~[red line in Fig.~\ref{fig3:quadrupole}(c)].
Although the sub-symmetry-protecting perturbation induces microscopic fluctuations in the local charge density, these modulations vanish upon coarse-graining over several unit cells, so that the macroscopic bulk charge density remains neutral.
In this coarse-grained limit, the charge missing from the non-quantized corner is effectively redistributed along the edges, and this boundary-localized redistribution is encoded in a non-quantized quadrupole moment.
Both the quadrupole moment of the corner state not protected by sub-symmetry and the bulk contribution $q_{xy}^\text{bulk}(m_1)$ remain non-quantized as a consequence of the charge redistribution.
By contrast, the sub-symmetry-protected corners maintain a constant, quantized $q^{i}_{xy}$ over the full range of $m_{1}$, evidencing robustness under the residual sub-symmetries~[blue line in Fig.~\ref{fig3:quadrupole}(c)].
Note that, as a result of chiral-symmetry breaking and the accompanying charge contribution, the total quadrupole $q^{\mathrm{tot}}_{xy}$ is no longer quantized.

\section{Second-Order Topological Semimetal}
\subsection{Model System}
\begin{figure*}[t]
\includegraphics[width=\textwidth]{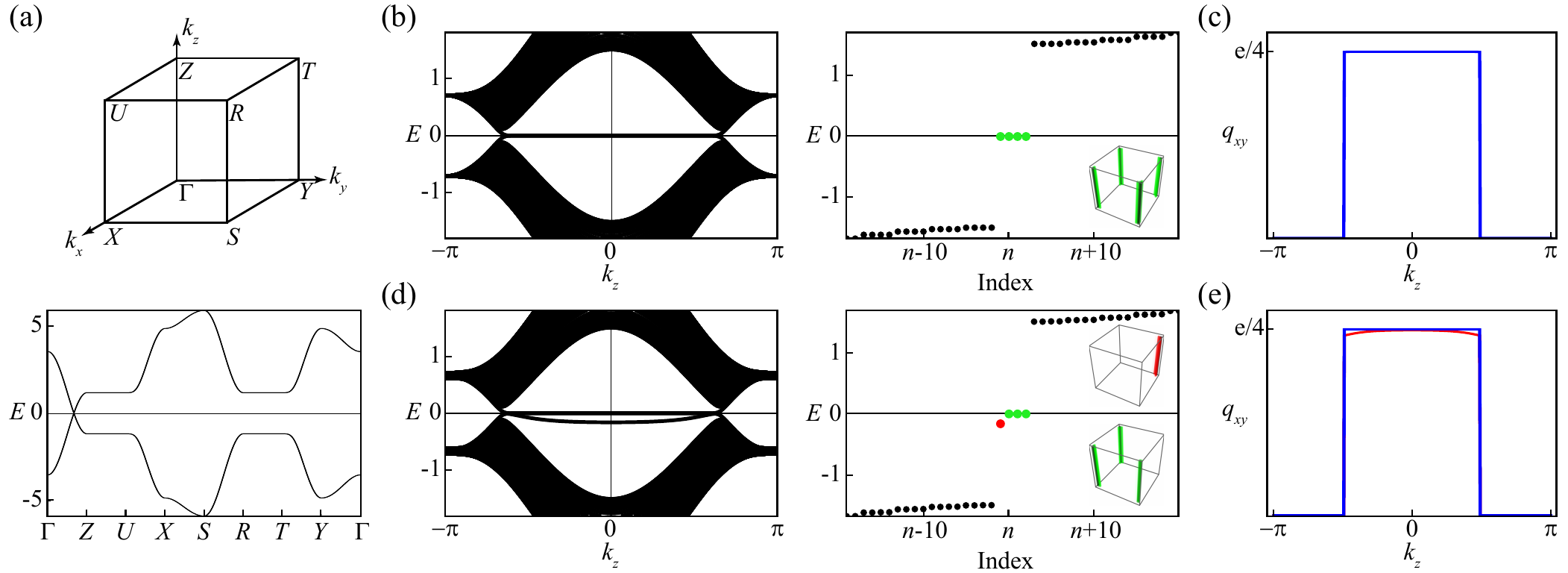}
\caption{\label{fig2:SOTSM}
\textbf{Modification of the second-order Fermi arcs in a SOTSM under sub-symmetry-protecting perturbation.}
\textbf{(a)} Three-dimensional Brillouin zone with labeled time-reversal-invariant momenta (TRIM) and the bulk band structure for parameters $\gamma = -0.5$, $\lambda = 1$, and $\lambda_z = 1$. Dirac cones appear along the $\Gamma Z$ high-symmetry line.
\textbf{(b)} (Left) Band structure in rod geometry, with a system finite in the $xy$ plane ($20 \times 20$ unit cells) and periodic along $z$, without sub-symmetry-protecting perturbation ($m_1 = 0$). Band dispersion shows fourfold-degenerate flat-band second-order Fermi arcs at zero energy. 
(Right) Energy spectrum at $k_z = 0$, highlighting four symmetry-protected hinge-localized zero-energy second-order Fermi arc states (green). The inset shows their corresponding spatially localized wavefunctions. The total number of states at $k_z = 0$ is $2n = 1600$.
\textbf{(c)} State-resolved quadrupole moments $q_{xy}$ of the hinge states as a function of $k_z$. All second-order Fermi arc states exhibit quantized values corresponding to a fractional charge of $e/4$.
\textbf{(d)} Same as panel (b), but with a finite sub-symmetry-protecting perturbation ($m_1 = 0.2$). 
One hinge-localized second-order Fermi arc state (red) is shifted away from zero energy due to the perturbation, while the other three (green) remain protected.
Insets show the spatial wavefunction distributions of the perturbed and unperturbed second-order Fermi arc states.
\textbf{(e)} State-resolved quadrupole moments $q_{xy}$ of the hinge states as a function of $k_z$ under perturbation. The sub-symmetry-protected second-order Fermi arcs (blue) retain quantized quadrupole values ($e/4$), while the perturbed second-order Fermi arc state (red) becomes non-quantized, indicating topological degradation under sub-symmetry-breaking perturbation.
}
\end{figure*}

So far, we have focused on SOTIs and examined how their boundary states respond to sub-symmetry-protecting perturbations. We now extend our analysis to topological semimetals, and investigate how Fermi arcs---specifically, the second-order Fermi arcs in a SOTSM---are affected by such perturbations. This enables us to investigate the applicability of sub-symmetry protection in gapless systems and to explore how the interplay between bulk topology and symmetry manifests in semimetallic phases.

Dirac semimetals can be realized by stacking layers of two-dimensional topological insulators~\cite{liu2014discovery,borisenko2014experimental,neupane2014observation,ganeshan2015constructing}. This stacking approach has also been applied to SOTSMs, where three-dimensional SOTSM phases emerge from layered two-dimensional SOTIs~\cite{wang2021engineering,wieder2020strong,chen2023quasicrystalline}.
Following this idea, we adopt a model of a SOTSM, constructed by stacking copies of the BBH model---introduced in the previous section---along the $z$-axis~\cite{wang2021engineering}.
In this construction, the interlayer coupling is introduced via the replacement $\lambda + \lambda_z \cos k_z$ for $\lambda$ in the Bloch Hamiltonian of Eq.~(\ref{eq:QI}), where $\lambda_z$ denotes the interplane hopping amplitude. 
This leads to the following three-dimensional Bloch Hamiltonian:
\begin{eqnarray}
\label{eq:SOTSM}
    H_{\text{SOTSM}}(\textbf{k}) =&& \left[ \gamma+\left(\lambda+\lambda_z \cos k_z\right) \cos k_x \right] \tau_x \\
&&-\left[ \gamma+\left(\lambda+\lambda_z \cos k_z\right) \cos k_x \right]\tau_y \sigma_y \nonumber\\ 
&&-\left(\lambda+\lambda_z \cos k_z\right) \sin k_x \tau_y \sigma_z\nonumber\\
&&-\left(\lambda+\lambda_z \cos k_z\right) \sin k_y \tau_y\sigma_x.\nonumber
\end{eqnarray}
The resulting Hamiltonian hosts bulk Dirac points along the $\Gamma Z$ line, as shown in Fig.~\ref{fig2:SOTSM}(a).
The corresponding second-order Fermi arcs are illustrated in Fig.~\ref{fig2:SOTSM}(b), confirming that the system realizes a SOTSM phase.
The model is of the BDI symmetry class, preserving time-reversal ($T=K$), particle-hole ($C=\tau_z K$), and chiral ($\Gamma=\tau_z$) symmetries inherited from the BBH Hamiltonian in Eq.~(\ref{eq:QI}).
These symmetries ensure the presence of zero-energy second-order Fermi arcs protected by chiral symmetry.
The schematic of the system is provided in Fig.~\ref{fig0:Schematics}(c), which shows four hinge-localized second-order Fermi arc states connecting the bulk Dirac points along the $\Gamma Z$ direction.

\subsection{Numerical and Analytical Analysis}
The topology-protecting symmetry in the SOTSM directly inherits that of the BBH model. Accordingly, we employ a perturbation analogous to that used in the BBH model.
For simplicity, we introduce a quasi-periodic on-site potential similar to the one used in the previous section:
\begin{eqnarray}
    m(i,j) = m_1 \cos\left[\pi\left(\sqrt{5}-1\right)(i+j)\right],
\end{eqnarray}
which is applied to the sublattice $1$ site of each unit cell.
Here, $i$ and $j$ indicate the position of the unit cell along the $x$ and $y$ directions, respectively.
This on-site potential is equally distributed for every sheet of the BBH model.

Figure~\ref{fig2:SOTSM}(d) presents the energy spectrum and corresponding wavefunction distribution for the SOTSM under a sub-symmetry-protecting perturbation ($m_1\neq0$).
The energy spectrum and corresponding wavefunction distribution reveal that the second-order Fermi arc associated with the hinge passing through the sublattice-1 site is shifted away from zero energy, a direct consequence of the chiral symmetry breaking via the perturbation.
Notably, despite this energy shift, the hinge state along the sublattice-1 site continues to connect the bulk Dirac points, just as the three sub-symmetry-protected second-order Fermi arcs do.
This suggests that the bulk Dirac nodes remain intact under the applied sub-symmetry-protecting perturbation, which primarily alters boundary-localized features while preserving the essential bulk band topology of this semimetal.
Small deformations of the bulk band structure may arise due to the sub-symmetry-protecting perturbation; however, the gapless Dirac nodes remain robust owing to symmetry constraints of the bulk semimetal.

For the analytical treatment of the SOTSM, we follow the same methodology as in the BBH model, with the key modification $\lambda+\lambda_z \cos k_z$ to incorporate interlayer couplings.
This substitution leads to a Dirac-type Hamiltonian,$H_{\text{SOTSM}}^\text{Dirac}$, analogous to Eq.~(\ref{eq:DiracBBH}), allowing the analytical form of the second-order Fermi arc states to be derived explicitly:
\begin{eqnarray}
\label{eq:DiracSOTSM}
    H_{\text{SOTSM}}^\text{Dirac} &=& i (\lambda+\lambda_z \cos k_z) \partial_r \tau_y\left( \cos\phi \sigma_z+\sin\phi\sigma_x\right)\\
&&+\alpha \left(\tau_x-\tau_y \sigma_y\right)+m(r) \left(\tau_0+\tau_z\right)\left(\sigma_0+\sigma_z\right)\nonumber.
\end{eqnarray}
Here, $m(r)$ is the perturbation term, corresponding to the continuum-limit approximation of the quasi-periodic on-site potential.
We adopt a cylindrical geometry, which reflects the physical configuration of the system, as all hinges---and thus the second-order Fermi arcs---extend along the $z$-axis.
The angle $\phi$ again parametrizes the boundary orientation in the $xy$-plane. The resulting solutions preserve the Jackiw–Rebbi structure, modulated by the momentum-dependent interlayer coupling.

Regardless of the specific form of the sub-symmetry-protecting perturbation, the analytical Jackiw–Rebbi solutions for the topological zero-energy hinge states in the SOTSM are given by:
\begin{eqnarray}
    \psi_{\phi=\frac{5\pi}{4}}(r)=&\mathcal{N}e^{\frac{\sqrt{2}\alpha}{\lambda+\lambda_z \cos k_z} (r-R)}
	\begin{pmatrix}
	0\\
	1\\
	0\\
	0\\
	\end{pmatrix},\\
    \psi_{\phi=\frac{3\pi}{4}}(r)=&\mathcal{N}e^{\frac{\sqrt{2}\alpha}{\lambda+\lambda_z \cos k_z} (r-R)}
	\begin{pmatrix}
	0\\
	0\\
	1\\
	0\\
	\end{pmatrix},\\
    \psi_{\phi=\frac{7\pi}{4}}(r)=&\mathcal{N}e^{\frac{\sqrt{2}\alpha}{\lambda+\lambda_z \cos k_z} (r-R)}
	\begin{pmatrix}
	0\\
	0\\
	0\\
	1\\
	\end{pmatrix},
\end{eqnarray}
where $R$ is the distance from the center to the hinge, $0\leq r \leq R$, and $\mathcal{N}$ is a normalization factor, respectively.
These three solutions describe exponentially localized hinge modes at zero energy, with the decay rate modulated by the momentum-dependent interlayer coupling $\lambda+\lambda_z \cos k_z$.
They correspond to sub-symmetry-protected second-order Fermi arc states in the SOTSM phase, each localized at a specific hinge located at angular positions $\phi=\frac{5\pi}{4}, \frac{3\pi}{4}$, and $\phi=\frac{7\pi}{4}$, respectively. The spatial localization of each state is determined by the residual sub-symmetry preserved along the corresponding boundary.

The hinge state at angle $\phi=\frac{\pi}{4}$, where the relevant sub-symmetry is broken, can be obtained by approximating the perturbation as a constant, $m(r)=m_1$.
Under this simplification, an analytical expression for the non-topological hinge arc is given by
\begin{eqnarray}
\label{eq:BrokenHinge}
    \psi(r)=\mathcal{N}e^{\frac{\beta}{\lambda+\lambda_z \cos k_z} (r-R)}
	\begin{pmatrix}
	2\alpha+\sqrt{2}\beta\\
	0\\
	E-m_1\\
	E-m_1\\
	\end{pmatrix},
\end{eqnarray}
where $\beta=\sqrt{2\alpha^2+E(m_1-E)}$ and $\mathcal{N}$ is the normalization constant, respectively.

The prefactor $\lambda+\lambda_z \cos k_z$, appearing in the denominator of the exponential factor in the wavefunction, determines the location of the hinges along the $k_z$-axis, which can be seen in the schematics of Fig.~\ref{fig0:Schematics}(c).
When $\lambda+\lambda_z \cos k_z>0$, the wavefunction decays exponentially as $r$ goes to infinity, indicating that the state is localized at the hinge.
In contrast, if $ \lambda + \lambda_z \cos k_z\leq0$, the wavefunction diverges and, therefore, cannot exist.
Accordingly, hinge states appear only within a finite window of 
$k_z$ connecting the bulk Dirac nodes---an observation that is fully consistent with the tight-binding spectra in Fig.~\ref{fig2:SOTSM}(b) and (d).

\subsection{Topological Invariants}
Figure~\ref{fig2:SOTSM}(c) illustrates the quadrupole moment of the hinge states in the SOTSM.
As in the SOTI case, the topological character of boundary states can be diagnosed via their quadrupole moments. Topological second-order Fermi arc states exhibit quantized quadrupole moments~\cite{chen2023quasicrystalline}, as clearly demonstrated in Fig.~\ref{fig2:SOTSM}(c), where all second-order Fermi arcs exhibit quantized values within the range of $k_z$ connecting the Dirac points.
This quantization occurs when the condition 
$|\gamma|<|\lambda+\lambda_z \cos k_z|$ is satisfied—a criterion that, for fixed $k_z$, is analogous to the topological regime of the BBH model,  $|\gamma|<|\lambda|$.
For fixed values of $k_z$, the SOTSM behaves as a two-dimensional insulator, except at the critical points satisfying 
$|\gamma|=|\lambda+\lambda_z \cos k_z|$, which mark the location of the bulk Dirac points. Consequently, the surface projection of these Dirac points and the second-order Fermi arcs connecting them can be interpreted as a momentum-resolved stack of SOTI systems with an effective $\lambda$ parameter~\cite{wieder2020strong,wang2021engineering}. Each layer follows the same topological criteria as the BBH model discussed earlier, leading to quantized quadrupole moments for the second-order Fermi arcs across the range of $k_z$, as confirmed in previous studies~\cite{wieder2020strong,chen2023quasicrystalline}.

By contrast, the application of sub-symmetry-protecting perturbation leads to a hinge state---specifically at $\phi=\frac{\pi}{4}$---in which the quadrupole moment is no longer quantized, as shown by the red curve in Fig.~\ref{fig2:SOTSM}(e).
This breakdown of quantization can be understood analytically by inserting the wavefunction $\psi(r)$ from Eq.~(\ref{eq:BrokenHinge}) into the expression for the quadrupole moment in Eq.~(\ref{eq:quad}), confirming the absence of topological protection along this perturbed hinge.
This behavior is further corroborated by our numerical calculations, which show that the quadrupole moment progressively deviates from its quantized value as $m_1$ increases, similar to Fig.~\ref{fig3:quadrupole}(c).
These findings highlight that sub-symmetry-protected boundary states are highly sensitive to the local symmetry structure—especially on specific hinges—and that quantization is preserved only when the corresponding sub-symmetry is intact.

\section{Conclusion}
In this work, we have investigated the role of sub-symmetry in higher-order topological systems. We have shown that sub-symmetry-protected boundary states emerge along boundaries where the sub-symmetry is preserved, even in the absence of the full topology-protecting symmetry. 
In a second-order topological insulator, these sub-symmetry-protected corner states exhibit robust topological features, including state-resolved quantized quadrupole moments, which underscores their stability under symmetry-breaking perturbations.
Extending this framework to a second-order topological semimetal, we demonstrated that sub-symmetry-protecting perturbations deform the second-order Fermi arc states while leaving the bulk Dirac points intact.
Crucially, only the second-order Fermi arcs on symmetry-broken hinges lose their quantized quadrupole moments, whereas those on sub-symmetry-protected hinges retain their topological character.
Furthermore, we have demonstrated that sub-symmetry-protected boundary states are not confined to high-symmetry corners but can also emerge along generic boundaries, as long as the corresponding sub-symmetry remains intact. 
This reinforces the idea that, like conventional higher-order boundary modes, sub-symmetry-protected states are fundamentally geometry-independent and dictated solely by underlying symmetry structure.

So far, our discussion has focused on systems in the BDI symmetry class, which possess time-reversal, particle-hole, and chiral symmetries. A natural direction for future research is to extend the sub-symmetry-protected framework to systems beyond this class—for instance, to the AI class, which preserves only time-reversal symmetry, or the A class, which lacks all three symmetries.
In particular, applying the sub-symmetry-protected concept to class A systems would shift the corresponding SOTSM from a Dirac semimetal to a Weyl semimetal. This change in symmetry class is expected to fundamentally alter the structure and connectivity of second-order Fermi arcs, potentially revealing new types of symmetry-selective topological boundary phenomena in gapless systems with reduced symmetry constraints.

Beyond theoretical modeling, the experimental realization of sub-symmetry-protected boundary states offers a potential direction for validating and extending our framework. Several platforms---including topolectrical circuits~\cite{ezawa2018higher,imhof2018topolectrical,lee2018topolectrical,dong2021topolectric} and photonic lattices~\cite{xie2018second,li2022higher,schulz2022photonic,el2019corner,zhou2024realization}---offer feasible routes to implementing the BBH model and its sub-symmetry-protected modifications.
Topological circuits are well-suited for implementing sub-symmetry-protecting perturbations, due to their modular architecture and fine-grained control over local parameters. 
The BBH model has already been realized in such systems using capacitors and inductors~\cite{imhof2018topolectrical,dong2021topolectric}.
A potential route toward realizing the sub-symmetry-protected BBH model in this platform involves sublattice-selective admittance modulations, with impedance spectroscopy serving as a viable probe of boundary-localized sub-symmetry-protected states.
Photonic systems have also successfully realized BBH-based higher-order topological insulators~\cite{xie2018second,li2022higher,schulz2022photonic,el2019corner,zhou2024realization}, and may engineer sub-symmetry-protected corner states and second-order Fermi arc modes via spatially modulated refractive indices or gain/loss profiles. 
While three-dimensional architectures remain challenging, topolectric and photonic systems may serve as conceptual analogs to our SOTSM construction.

Altogether, our findings highlight that topology, even in its higher-order and sub-symmetry-resolved forms, remains a robust organizing principle across both gapped and gapless systems.

\section*{Acknowledgments}
This work was supported by the National Research Foundation of Korea (NRF) funded by the Ministry of Science and ICT (MSIT), South Korea (Grants No. NRF-2022R1A2C1011646, RS-2024-00416036, and RS-2025-03392969).
This work was supported by Creation of the Quantum Information Science R\&D Ecosystem (Grant No. RS-2023-NR068116) through the NRF funded by the  Korean government (MSIT).
This work was supported by the Quantum Simulator Development Project for Materials Innovation through the NRF funded by the MSIT, South Korea (Grant No. RS-2023-NR119931).
This work was also supported by Brain Pool program funded by the Ministry of Science and ICT through the NRF (Grant No. RS-2025-25446099).
M. Kang, W. Sung, and S. Cheon also acknowledge support from the POSCO Science Fellowship of the POSCO TJ Park Foundation.

\appendix
\section{Effective Boundary Hamiltonian of SOTI}\label{Appendix:Surface}
In this section, we derive the edge Hamiltonian of the BBH model following the method outlined in Ref.~\cite{schindler2020dirac}.
We begin with the low-energy effective Hamiltonian in Eq.~(\ref{eq:QID}), which is 
\begin{eqnarray}
\label{eq:AQI}
    H_{\text{BBH}}^\text{eff}(k_\perp, k_\parallel)  &=& -\lambda k_\perp \tau_y\left( \cos\phi \sigma_z+\sin\phi\sigma_x\right)\\
&&+\lambda k_\parallel \tau_y\left(\sin\phi \sigma_z-\cos\phi\sigma_x\right)\nonumber\\
&&+\alpha \left(\tau_x-\tau_y \sigma_y\right)\nonumber.
\end{eqnarray}
We consider a circular boundary geometry in real space, where the perpendicular momentum acts as a radial differential operator $k_\perp=-i\partial_r$ in the radial direction, while the tangential momentum $k_\parallel$ remains a good quantum number~\cite{schindler2020dirac}.
For simplicity, we set $\lambda=1$ throughout the following derivation.

To proceed, we decompose the low-energy effective Hamiltonian into two components: a Dirac-type part that governs the localized modes on the boundary ($r=R$), and a perturbative part~\cite{schindler2020dirac}. The Dirac Hamiltonian is given by
\begin{eqnarray}
\label{eq:AQID}
	H_\text{D}&=&i\partial_r \tau_y\left( \cos\phi \sigma_z+\sin\phi\sigma_x\right)\\
    &&+\alpha \left(|\cos \phi|\tau_x-|\sin \phi|\tau_y \sigma_y\right),\nonumber
\end{eqnarray}
while the perturbation part reads
\begin{eqnarray}
\label{eq:AQIP}
	H_\text{per}&=&k_\parallel \tau_y\left(\sin\phi \sigma_z-\cos\phi\sigma_x\right)\\
    &&+\alpha \left[\left(1-|\cos \phi|\right)\tau_x-\left(1-|\sin \phi|\right)\tau_y \sigma_y\right].\nonumber
\end{eqnarray}
This decomposition is chosen such that the resulting expressions are consistent with the edge Hamiltonians obtained for the top, bottom, right, and left boundaries in Appendix~\ref{Appendix:Edge}. 
It further allows the Jackiw–Rebbi–type bound-state solutions to be factorized into radial and angular components. Due to the absolute-value terms in Eq.~\eqref{eq:AQID}, however, the edge spectrum must be analyzed separately in four angular sectors:
The split is along $0\leq\phi\leq\frac{\pi}{2}$, $\frac{\pi}{2}\leq\phi\leq\pi$, $\pi\leq\phi\leq\frac{3\pi}{2}$, and $\frac{3\pi}{2}\leq\phi\leq 2\pi$, which we denote as first, second, third, and fourth quarters, respectively.

Solving the Dirac equation $H_\text{D}\psi=0$, we obtain Jackiw–Rebbi–type solutions localized at the boundary $r=R$ for each quadrant $q=1,2,3,4$. The solutions take the form
\begin{eqnarray}
    \psi_{1,2}^q(r) = \mathcal{N}_{1,2}^q \, e^{\alpha (r-R)} \, u_{1,2}^q(\phi),
\end{eqnarray}
where $\mathcal{N}_{i}^q$ is a normalization constant, $u_i^q(\phi)$ denotes the angular spinor component for $i=1,2$, and $0 \leq r \leq R$. The angular parts are explicitly given by
\begin{eqnarray}
    u^1_1=\begin{pmatrix}
        1\\0\\0\\0
    \end{pmatrix},~
    u^1_2=\begin{pmatrix}
        0\\0\\\sin\phi\\-\cos\phi
    \end{pmatrix}, \\
    u^2_1=\begin{pmatrix}
        \sin\phi\\-\cos\phi\\0\\0
    \end{pmatrix},~
    u^2_2=\begin{pmatrix}
        0\\0\\1\\0
    \end{pmatrix}, \\
    u^3_1=\begin{pmatrix}
        0\\1\\0\\0
    \end{pmatrix},~
    u^3_2=\begin{pmatrix}
        0\\0\\-\cos\phi\\-\sin\phi
    \end{pmatrix}, \\
    u^4_1=\begin{pmatrix}
        \cos\phi\\\sin\phi\\0\\0
    \end{pmatrix},\quad
    u^4_2=\begin{pmatrix}
        0\\0\\0\\-1
    \end{pmatrix}.
\end{eqnarray}

Projecting the perturbative Hamiltonian $H_\text{per}$ onto the subspace spanned by these boundary-localized states yields an effective edge Hamiltonian for each quadrant,
\begin{eqnarray}
    H_{\text{BBH}}^{\text{edge},q}=\sum_{i,j}\bra{\psi_i^q} H_\text{per}^q \ket{\psi_j^q}.
\end{eqnarray}
Collecting contributions from all four quadrants, the total effective edge Hamiltonian is
\begin{eqnarray}
\label{eq:sotiedge}
    H_{\text{BBH}}^\text{edge} = k_\parallel \sigma_y + \alpha(\phi)\sigma_x,
\end{eqnarray}
where $\sigma_i$ are Pauli matrices acting in the reduced edge subspace. The angular-dependent Dirac mass takes the form
\begin{eqnarray}
    \alpha(\phi) = \pm \alpha \big( \cos\phi \mp \sin\phi \big),
\end{eqnarray}
with the sign pattern $(+,-)$, $(-,+)$, $(-,-)$, and $(+,+)$ for the first, second, third, and fourth quadrants, respectively.  

This effective edge Hamiltonian coincides with Eq.~(\ref{eq:H_QI_surface}) in the main text. Importantly, the Dirac mass $\alpha(\phi)$ undergoes a sign reversal at angular positions $\phi = \tfrac{\pi}{4},\, \tfrac{3\pi}{4},\, \tfrac{5\pi}{4},\, \tfrac{7\pi}{4}$, giving rise to domain walls and consequently zero-energy corner states, as illustrated in Fig.~\ref{fig4:DiracMass}(a).

When the sub-symmetry-protecting perturbation $m(r)$ is introduced on the sublattice-1 site, in analogy with Eq.~\eqref{eq:DiracBBH}, the perturbation part in Eq.~\eqref{eq:AQIP} acquires an additional contribution and becomes
\begin{eqnarray}
\label{eq:AQIP2}
	H_\text{per} &=& k_\parallel \tau_y \left( \sin\phi \,\sigma_z - \cos\phi \,\sigma_x \right) \\
    &&+ \alpha \left[ \big(1-|\cos \phi|\big)\tau_x - \big(1-|\sin \phi|\big)\tau_y \sigma_y \right] \nonumber \\
    &&+ \frac{m_1}{4} \left( \tau_0 + \tau_z \right)\left( \sigma_0 + \sigma_z \right). \nonumber
\end{eqnarray}
For analytical tractability, the perturbation has been approximated by a constant $m_1$. The Dirac Hamiltonian itself in \eqref{eq:AQID} remains unchanged, and thus the unperturbed edge-localized states are preserved. The resulting total effective edge Hamiltonian reads
\begin{eqnarray}
\label{eq:sotiedge2}
    H_{\text{BBH}}^\text{edge} = k_\parallel \sigma_y + \alpha(\phi)\sigma_x + m(\phi)\,\frac{\sigma_0+\sigma_z}{2},
\end{eqnarray}
where the angular dependence of the perturbation is piecewise defined as
\begin{eqnarray}
\label{eq:sotiedge2mphi}
m(\phi) =
\begin{cases}
m_1, & 0 \leq \phi \leq \pi/2, \\
m_1 \sin^2 \phi, & \pi/2 \leq \phi \leq \pi, \\
0, & \pi \leq \phi \leq 3\pi/2, \\
m_1 \cos^2 \phi, & 3\pi/2 \leq \phi \leq 2\pi.
\end{cases}
\end{eqnarray}
This quadrant dependence reflects the fact that the perturbation acts only on sublattice-1, which contributes to the boundary theory in the first, second, and fourth quadrants, but is absent in the third.

The domain-wall (corner) modes can be obtained in the continuum limit by linearizing the edge theory around each corner 
$\phi_0 \in \{\tfrac{\pi}{4},\,\tfrac{3\pi}{4},\,\tfrac{5\pi}{4},\,\tfrac{7\pi}{4}\}$.
Defining the local angular deviation $\bar\phi$ by $\phi=\phi_0+\bar\phi$, the linearized Hamiltonians take the form
\begin{eqnarray}
    H_{\text{JR}}^{\text{edge},1} &=& -i\,\partial_{\bar{\phi}}\,\sigma_y
    + \alpha\,\bar{\phi}\,\sigma_x
    + m(\bar{\phi})\,\frac{\sigma_0+\sigma_z}{2}, \label{eq:JR_edge_1}\\
    H_{\text{JR}}^{\text{edge},2} &=& -i\,\partial_{\bar{\phi}}\,\sigma_y
    - \alpha\,\bar{\phi}\,\sigma_x
    + m(\bar{\phi})\,\frac{\sigma_0+\sigma_z}{2},\label{eq:JR_edge_2}
\end{eqnarray}
where $H_{\text{JR}}^{\text{edge},1}$ applies to the first and third quadrants, and 
$H_{\text{JR}}^{\text{edge},2}$ to the second and fourth. 
The angular dependent constant term $m(\bar\phi)$ is given piecewise by Eq.~\eqref{eq:sotiedge2mphi}.

Setting $m(\bar\phi)=0$, the zero-energy Jackiw--Rebbi equations 
$H_{\text{JR}}^{\text{edge},a}\psi^{(a)}_0=0$ ($a=1,2$) decouple into first-order ordinary differential equations for the spinor components. 
Each yields a single normalizable zero mode localized at the corner, which we write as
\begin{eqnarray}
    \psi^{(1)}_0(\bar\phi) &=& \mathcal{N}_1\,e^{\alpha \bar{\phi}}(\bar\phi)\,
    \begin{pmatrix} 1 \\ 0 \end{pmatrix}, 
    \label{eq:JR_zeromodes1}
    \\
    \psi^{(2)}_0(\bar\phi) &=& \mathcal{N}_2\,e^{-\alpha \bar{\phi}}(\bar\phi)\,
    \begin{pmatrix} 0 \\ 1 \end{pmatrix},
    \label{eq:JR_zeromodes2}
\end{eqnarray}
with $\mathcal{N}_{1,2}$ normalization constants, and $e^{\alpha \bar{\phi}}$ and $e^{-\alpha \bar{\phi}}$ indicate decay away from the corner. 
These two spinors form the local edge subspace on which we project the perturbations.

Reinstating the sub-symmetry-protecting term, 
$H^{\text{edge}}_m(\bar\phi)=m(\bar\phi)\,(\sigma_0+\sigma_z)/2$, the leading energy corrections are obtained by projection:
\begin{eqnarray}
\label{eq:JR_energies}
    E_i \;=\; \big\langle \psi^{(i)}_0 \big| H^{\text{edge}}_m \big| \psi^{(i)}_0 \big\rangle, 
    \qquad i=1,2,
\end{eqnarray}
with off-diagonal matrix elements vanishing due to the orthogonality of the spinors in Eqs.~\eqref{eq:JR_zeromodes1} and \eqref{eq:JR_zeromodes2}. 
As $H^{\text{edge}}_m$ projects onto the $(\sigma_0+\sigma_z)$ sector, only the mode with support on the upper spinor component (the $\sigma_z=+1$ component) acquires a finite shift. 
Consequently,
\begin{eqnarray}
    E_1 &=& m(\bar \phi),
    \\
    E_2 &=& 0,
\end{eqnarray}
and energy correction is nonzero only for the first and third quadrants.
Using the quadrant-dependent $m(\phi)$ from Eq.~\eqref{eq:sotiedge2mphi}, one finds that only the corner at 
$\phi_0=\tfrac{\pi}{4}$ (first quadrant) exhibits a nonzero shift, while the other three corners remain pinned at zero energy. 
Equivalently, the sub-symmetry-protecting perturbation term lifts exactly one corner mode---leaving three sub-symmetry-protected zero modes---in agreement with Fig.~\ref{fig4:DiracMass}(b) and with the one-dimensional sub-symmetey-protecting mechanism discussed in Ref.~\cite{kang2024subsymmetry}. 
Although the approach seems perturbative, it should be noted that the solution of Eqs.~\eqref{eq:JR_zeromodes1} and \eqref{eq:JR_zeromodes2} are exact solutions of the Hamiltonians in Eqs.~\eqref{eq:JR_edge_1} and \eqref{eq:JR_edge_2}.
A complementary edge-by-edge derivation on a square geometry is provided in Appendix~\ref{Appendix:Edge}.

\section{Effective Edge and Surface Hamiltonian Analysis}\label{Appendix:Edge}

In this section, we derive the effective Hamiltonians for each edge of the BBH model from a one-dimensional perspective, following the method introduced in Ref.~\cite{schindler2020dirac}. We start from the Bloch Hamiltonian given in Eq.~(\ref{eq:QI}), reproduced here for convenience:
\begin{eqnarray}
    H_{\text{BBH}}(\mathbf{k}) &=& \left( \gamma + \lambda \cos k_x \right) \tau_x - \lambda \sin k_x \tau_y \sigma_z \\
    && - \left( \gamma + \lambda \cos k_y \right) \tau_y \sigma_y - \lambda \sin k_y \tau_y \sigma_x. \nonumber
\end{eqnarray}
We consider a square-shaped geometry in real space. For the left and right edges, the system is finite along the $x$-direction and translationally invariant along $y$, so we replace $k_x = -i \partial_x$ while keeping $k_y$ as a good quantum number. Similarly, for the top and bottom edges, we set $k_y =-i \partial_y$ and treat $k_x$ as a parameter.
Throughout the derivation, we set $\lambda = 1$ for simplicity.

To proceed, we decompose the low-energy effective Hamiltonian into two components: a Dirac-like term that determines the edge-localized modes and a perturbative term~\cite{schindler2020dirac}. We define $\alpha \equiv \gamma + \lambda$, and work in the low-energy limit near the $\Gamma$ point.
For simplicity, we also approximate the perturbation of Eq.~(\ref{eq:SOTIpert}) as $m_1$.

For the top and bottom edges, where the system is finite along $y$, the Dirac-like Hamiltonian and the perturbation term are given by
\begin{eqnarray}
    H_\text{D}^{\text{top/bottom}} &=& i\partial_{y} \, \tau_y \sigma_x - \alpha \, \tau_y \sigma_y, \\
    H_\text{per}^{\text{top/bottom}} &=& \left( \gamma + \cos k_x \right) \tau_x - \sin k_x \, \tau_y \sigma_z \nonumber\\
    &&+ \frac{m_1}{4} \left( \tau_0 + \tau_z \right) \left( \sigma_0 + \sigma_z \right),
\end{eqnarray}
where $m_1$ denotes the strength of the sub-symmetry-protecting perturbation.

Similarly, for the left and right edges, where the system is finite along $x$, the Dirac-like and perturbation components are given by
\begin{eqnarray}
    H_\text{D}^{\text{left/right}} &=& i\partial_{x} \, \tau_y \sigma_z + \alpha \, \tau_x, \\
    H_\text{per}^{\text{left/right}} &=& - \left( \gamma + \cos k_y \right) \tau_y \sigma_y - \sin k_y \, \tau_y \sigma_x \nonumber\\
    &&+ \frac{m_1}{4} \left( \tau_0 + \tau_z \right) \left( \sigma_0 + \sigma_z \right).
\end{eqnarray}

Following the projection methodology developed in Appendix~\ref{Appendix:Surface}, we obtain the effective Hamiltonians for each edge of the BBH model in terms of Pauli matrices acting on the edge subspace.
The effective Hamiltonians for the four edges will be,
\begin{eqnarray}
	H_\text{eff}^\text{top}&=&\left( \gamma+ \cos k_x \right) \sigma_x+\sin k_x \sigma_y\label{eq:QIET}\\
    &&+\frac{m_1}{2}\left(\sigma_0+\sigma_z\right),\nonumber\\
    H_\text{eff}^\text{bottom}&=&\left( \gamma+ \cos k_x \right) \sigma_x+\sin k_x \sigma_y,\label{eq:QIEB}\\
    H_\text{eff}^\text{left}&=&-\left( \gamma+\cos k_y\right)\sigma_x+\sin k_y \sigma_y\label{eq:QIEL}\\
       H_\text{eff}^\text{right}&=&-\left( \gamma+\cos k_y\right)\sigma_x+\sin k_y \sigma_y\label{eq:QIER}\\
    &&+\frac{m_1}{2}\left(\sigma_0+\sigma_z\right).\nonumber
\end{eqnarray}

\begin{figure}[t]
\includegraphics[width=0.5\textwidth]{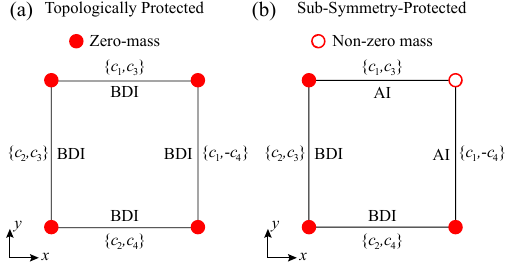}
\caption{\label{fig4:DiracMass}
\textbf{Edge symmetry classification, sublattice basis, and corner Dirac mass structure.}
\textbf{(a)} Without sub-symmetry-protecting perturbation, each edge belongs to the BDI class, with its Hamiltonian defined in the two-component sublattice basis. 
All four corners host zero Dirac mass (filled circles).
\textbf{(b)} Under the sub-symmetry-protecting perturbation, the edges involving sublattice-1 change their classification from BDI to AI. 
Consequently, only the corresponding corners acquire a finite Dirac mass (open circle), while the other corners remain massless and topologically protected (filled circles).  
}
\end{figure}

For each boundary, we work in a reduced two-component sublattice basis
$
\Psi^a(k)
=
\bigl(c_{i}(k),\,c_{j}(k)\bigr)^{T}
$,
where $k$ is a momentum conserved along the edge $a\in\{\text{top},\text{bottom},\text{left},\text{right}\}$.
The operators $c_{i}(k)$ annihilate a fermion on sublattice $i=1,2,3,4$ with momentum $k$.
For convenience, the edge spinors are chosen as
\begin{eqnarray}
    \Psi^{\text{top}}(k) &=&
    \begin{pmatrix}
    c_{1}(k)\\[2pt]
    c_{3}(k)
    \end{pmatrix},
    \\
    \Psi^{\text{bottom}}(k) &=&
    \begin{pmatrix}
    c_{2}(k)\\[2pt]
    c_{4}(k)
    \end{pmatrix},
    \\
    \Psi^{\text{left}}(k) &=&
    \begin{pmatrix}
    c_{2}(k)\\[2pt]
    c_{3}(k)
    \end{pmatrix},\\
    \Psi^{\text{right}}(k)  &=&
    \begin{pmatrix}
    c_{1}(k)\\[2pt]
    -c_{4}(k)
    \end{pmatrix}.
\end{eqnarray}
On this basis, the Pauli matrices $\sigma_{i}$ of the edge Hamiltonians in Eqs.~(\ref{eq:QIET}-\ref{eq:QIER}) act on these two-component sublattice spaces. 

The sub-symmetry-protecting perturbation contributes only to the top and right edges, as reflected by the last terms in Eqs.~(\ref{eq:QIET}) and (\ref{eq:QIER}).

When the sub-symmetry-protecting perturbation is absent ($ m_1 = 0 $), the edge Hamiltonians exhibit time-reversal symmetry $ T^a = K $, particle-hole symmetry $ C^a = \sigma_z K $, and chiral symmetry $ \Gamma^a = \sigma_z $ for the edges $a\in$ \{top, bottom, left, right\}.
As a result, the edges fall into the BDI symmetry class~\cite{schnyder2008classification,chiu2016classification}, which supports an integer topological invariant—namely, the winding number in one dimension.
In contrast, when the perturbation is present ($ m_1 \neq 0 $), the effective Hamiltonians for the top and right edges [Eqs.~(\ref{eq:QIET}) and (\ref{eq:QIER})] no longer possess particle-hole or chiral symmetry, although time-reversal symmetry $ T^\alpha = K $ remains intact. Consequently, these perturbed edges belong to the AI symmetry class~\cite{schnyder2008classification,chiu2016classification}, for which no topological invariant exists in one dimension.
Table~\ref{table1} summarizes, for each edge under the sub-symmetry-protecting perturbation, the presence or absence of time-reversal ($T$), particle–hole ($C$), and chiral ($\Gamma$) symmetries and the resulting Altland–Zirnbauer class~[see also Fig.~\ref{fig4:DiracMass}].

To analyze the impact of the sub-symmetry-protecting perturbation  on the boundary theory, 
we consider the linearized edge Hamiltonians near the $\Gamma$ point. 
For each edge $a \in \{\text{top},\text{bottom},\text{left},\text{right}\}$, 
the low-energy effective Hamiltonian takes the form
\begin{eqnarray}
    H_\text{lin}^a = k_i \sigma_y + m^a_0 \sigma_x + m^a_1 \left(\sigma_0 + \sigma_z\right),
    \label{eq:LinHam}
\end{eqnarray}
where $k_i = k_x$ for $a \in \{\text{top},\text{bottom}\}$ and 
$k_i = k_y$ for $a \in \{\text{left},\text{right}\}$. 
Here, $m^a_0 \sigma_x$ denotes the conventional Dirac mass term, 
while $m^a_1$ represent additional symmetry-breaking Dirac mass contributions induced by sub-symmetry-protecting perturbation.
It should be noted that this low-energy effective Hamiltonian of the edges is equal to the Su-Schrieffer-Heeger chain under sub-symmetry-protecting perturbation, which has been discussed in previous studies~\cite{kang2024subsymmetry}.
Based on this low-energy effective Hamiltonian and the edge Hamiltonians of Eqs.~(\ref{eq:QIET}-\ref{eq:QIER}), the Dirac mass term on the corresponding two-component sublattice spaces can be obtained as $m^\alpha = m^a_0\sigma_x+m^a_1 \left(\sigma_0 + \sigma_z\right)$.

These effective Dirac mass terms $m^\alpha$ can be directly evaluated using the Jackiw-Rebbi zero-mode solutions~\cite{kang2024subsymmetry}, 
$\Psi^\alpha_{1,2}$, which are localized along the edges in the corresponding two-component sublattice spaces: 
\begin{eqnarray}
    \Psi^a_1 \propto \begin{pmatrix} 1 \\ 0 \end{pmatrix},~~
    \Psi^a_2 \propto \begin{pmatrix} 0 \\ 1 \end{pmatrix}.
\end{eqnarray}
The corner effective Dirac masses are then obtained as
\begin{eqnarray}
    M^a_k = \langle \Psi^a_k | m^a | \Psi^a_k \rangle,
    \label{eq:CornerDirac}
\end{eqnarray}
where $k=1,2$.

When the sub-symmetry-protecting perturbation is applied to sublattice-1, the terms $m^\alpha_1 \sigma_0$ and $m^\alpha_2 \sigma_z$ 
act selectively on boundary states localized at this sublattice. 
Since the unperturbed Dirac mass $m^\alpha_0$ vanishes at the corners, the effective mass contribution can be directly evaluated using the Jackiw--Rebbi zero-mode solutions, 
$\Psi^\alpha_{1,2}$, which are localized along the edges in the corresponding two-component sublattice spaces: 
\begin{eqnarray}
    \Psi^\alpha_1 \propto \begin{pmatrix} 1 \\ 0 \end{pmatrix},~~
    \Psi^\alpha_2 \propto \begin{pmatrix} 0 \\ 1 \end{pmatrix}.
\end{eqnarray}
The corner effective Dirac masses are then obtained as
\begin{eqnarray}
    M^\alpha_k = \langle \Psi^\alpha_k | m^\alpha | \Psi^\alpha_k \rangle, \qquad k=1,2,
    \label{eq:CornerDirac}
\end{eqnarray}
where $m^\alpha = m^\alpha_1\sigma_0+m^\alpha_2\sigma_z$.

As an explicit example, for the top edge, the perturbation reduces to
\begin{eqnarray}
    m^{\text{top}} = -\frac{m_1}{2}\,(\sigma_0+\sigma_z)
    = \begin{pmatrix} -m_1 & 0 \\ 0 & 0 \end{pmatrix},
\end{eqnarray}
so that
$M^{\text{top}}_1 \propto \langle \Psi^{\text{top}}_1 | m^{\text{top}} | \Psi^{\text{top}}_1 \rangle = -m_1, 
~
M^{\text{top}}_2 = 0$.
Thus, only the state localized on sublattice-1 acquires a finite Dirac mass. 
A similar calculation for the right edge shows that $\Psi^{\text{right}}_2$ acquires a finite mass, 
while $\Psi^{\text{right}}_1$ remains massless. 
For the left and bottom edges, both modes yield vanishing expectation values. 
Consequently, only these specific corners develop a finite Dirac mass and lose their zero-energy character, 
whereas the remaining corners remain massless and topologically protected~[Fig.~\ref{fig4:DiracMass}(b)].

From the viewpoint of topological classification, the presence of chiral symmetry allows for the definition of a chiral winding number $\nu$ as a topological invariant for systems in the BDI class~\cite{schnyder2008classification,chiu2016classification}. 
For the edge Hamiltonians derived in Eqs.~(\ref{eq:QIET})–(\ref{eq:QIER}), the winding number is nontrivial in the topological regime $ |\gamma| < 1 $, and is computed via the standard expression:
\begin{eqnarray}
    \nu &=& \frac{1}{2\pi i} \int dk \, \partial_k \ln q(k) \\
        &=& \frac{1}{2\pi i} \int dk \, \partial_k \ln \left[\pm\left( \gamma + \cos k \right) - i \sin k \right] = 1,\nonumber
\end{eqnarray}
where $ q(k) $ is the off-diagonal component (the so-called $ q $-matrix) of the flattened Hamiltonian in the chiral basis, and $ k $ denotes the crystal momentum along the edge. The $ + $ and $ - $ signs correspond to the top/right and bottom/left edges, respectively. This result confirms that, in the absence of the sub-symmetry-breaking perturbation ($ m = 0 $), all four edges host nontrivial one-dimensional topological phases characterized by $ \nu = 1 $.

\begin{table}[t]
\begin{tabular}{c c c c c c c} 
 \hline
 ~~Surface~~& ~~~~~~~~~$T$~~~~~~~~~ & ~~~~~~~~~$C$~~~~~~~~~ & ~~~~~~~~~$\Gamma$~~~~~~~~~ &~~~~~~~~~Class~~~~~~~~~\\
 \hline
back & 1 & 0 & 0 & AI\\ 
front & 1 & 1 & 1 & BDI\\
left & 1 & 1 & 1 & BDI\\
right & 1 & 0 & 0 & AI\\
 \hline
\end{tabular}
\caption{
\textbf{Symmetry content and Altland-Zirnbauer classes of surface effective Hamiltonians under a sub-symmetry-protecting perturbation.}
Entries indicate presence (1) or absence (0) of time-reversal $T$, particle--hole $C$, and chiral $\Gamma$ symmetries; for $T$ and $C$, $1$ denotes $T^{2}=+1$, $C^{2}=+1$.
The resulting Altland-Zirnbauer class for each edge is listed in the last column.
This results in the second-order Fermi arc created by the back and right edge to have a non-quantized quadrupole moment.
}
\label{table2}
\end{table}

The introduction of the sub-symmetry-protecting perturbation explicitly breaks the chiral symmetry on the affected edges. As a result, the chiral winding number becomes ill-defined, and no conventional topological invariant can be assigned.
However, as discussed in the main text, boundary-localized states can still emerge at edges where the relevant sub-symmetry is preserved, despite the absence of a global topological index. These residual modes, which arise from selective symmetry protection, constitute the sub-symmetry-protected boundary states and reflect the underlying local symmetry structure of the system.

For the SOTSM, the bulk system has four surfaces (front, back, left, and right surfaces).
The effective surface Hamiltonians can be obtained from the effective edge Hamiltonians in Eqs.~(\ref{eq:QIET})-(\ref{eq:QIER}) as the SOTSM is generated from the SOTI system.
Then, the effective surface Hamiltonians are given by
\begin{eqnarray}
	H_\text{eff}^\text{back}&=&\left( \gamma+ \Tilde{\lambda}\cos k_x \right) \sigma_x+\Tilde{\lambda}\sin k_x \sigma_y\\
    &&+\frac{m_1}{2}\left(\sigma_0+\sigma_z\right),\nonumber\\
    H_\text{eff}^\text{front}&=&\left( \gamma+\Tilde{\lambda} \cos k_x \right) \sigma_x+\Tilde{\lambda}\sin k_x \sigma_y,\\
    H_\text{eff}^\text{left}&=&-\left( \gamma+\Tilde{\lambda}\cos k_y\right)\sigma_x+\Tilde{\lambda}\sin k_y \sigma_y\\
       H_\text{eff}^\text{right}&=&-\left( \gamma+\Tilde{\lambda}\cos k_y\right)\sigma_x+\Tilde{\lambda}\sin k_y \sigma_y\\
    &&+\frac{m_1}{2}\left(\sigma_0-\sigma_z\right),\nonumber
\end{eqnarray}
where $\Tilde{\lambda}=\lambda+\lambda_z \cos k_z$ and $\sigma_{x,y,z}$ act on the two–component surface sublattice basis inherited from the edge basis.
The $\sigma_0$ pieces are chemical–potential shifts; the $\sigma_z$ pieces commute with $\Gamma=\sigma_z$ and therefore break chiral symmetry on the affected surfaces.

Each surface inherits the Altland-Zirnbauer class of its parent two-dimensional edge Hamiltonians.
The effective surface Hamiltonians come from the effective edge Hamiltonians:
The top (bottom) edge Hamiltonian of the SOTI corresponds to the back (front) surface Hamiltonian of the SOTSM.
The left and right surface Hamiltonians of the SOTSM come from the corresponding edge Hamiltonian.
Therefore, the symmetry analysis in Table~\ref{table1} carries over layer by layer to the stacked construction of the SOTSM as shown in Table~\ref{table2}.

Under the sub-symmetry-protecting perturbation, the back and right surface systems (AI class) lose chiral symmetry, and the second-order Fermi arc between the two becomes dispersive.
On the other hand, the other surface systems are of the BDI class, and their second-order Fermi arc states remain topological~\cite{chiu2016classification}. 
This accounts for the observation that exactly one arc is affected while the other three remain quantized and robust, as can be seen in Fig.~\ref{fig2:SOTSM}.


\begin{thebibliography}{40}%
\makeatletter
\providecommand \@ifxundefined [1]{%
 \@ifx{#1\undefined}
}%
\providecommand \@ifnum [1]{%
 \ifnum #1\expandafter \@firstoftwo
 \else \expandafter \@secondoftwo
 \fi
}%
\providecommand \@ifx [1]{%
 \ifx #1\expandafter \@firstoftwo
 \else \expandafter \@secondoftwo
 \fi
}%
\providecommand \natexlab [1]{#1}%
\providecommand \enquote  [1]{``#1''}%
\providecommand \bibnamefont  [1]{#1}%
\providecommand \bibfnamefont [1]{#1}%
\providecommand \citenamefont [1]{#1}%
\providecommand \href@noop [0]{\@secondoftwo}%
\providecommand \href [0]{\begingroup \@sanitize@url \@href}%
\providecommand \@href[1]{\@@startlink{#1}\@@href}%
\providecommand \@@href[1]{\endgroup#1\@@endlink}%
\providecommand \@sanitize@url [0]{\catcode `\\12\catcode `\$12\catcode `\&12\catcode `\#12\catcode `\^12\catcode `\_12\catcode `\%12\relax}%
\providecommand \@@startlink[1]{}%
\providecommand \@@endlink[0]{}%
\providecommand \url  [0]{\begingroup\@sanitize@url \@url }%
\providecommand \@url [1]{\endgroup\@href {#1}{\urlprefix }}%
\providecommand \urlprefix  [0]{URL }%
\providecommand \Eprint [0]{\href }%
\providecommand \doibase [0]{http://dx.doi.org/}%
\providecommand \selectlanguage [0]{\@gobble}%
\providecommand \bibinfo  [0]{\@secondoftwo}%
\providecommand \bibfield  [0]{\@secondoftwo}%
\providecommand \translation [1]{[#1]}%
\providecommand \BibitemOpen [0]{}%
\providecommand \bibitemStop [0]{}%
\providecommand \bibitemNoStop [0]{.\EOS\space}%
\providecommand \EOS [0]{\spacefactor3000\relax}%
\providecommand \BibitemShut  [1]{\csname bibitem#1\endcsname}%
\let\auto@bib@innerbib\@empty
\bibitem [{\citenamefont {Hasan}\ and\ \citenamefont {Kane}(2010)}]{hasan2010colloquium}%
  \BibitemOpen
  \bibfield  {author} {\bibinfo {author} {\bibfnamefont {M~Zahid}\ \bibnamefont {Hasan}}\ and\ \bibinfo {author} {\bibfnamefont {Charles~L}\ \bibnamefont {Kane}},\ }\bibfield  {title} {\enquote {\bibinfo {title} {Colloquium: topological insulators},}\ }\href@noop {} {\bibfield  {journal} {\bibinfo  {journal} {Reviews of modern physics}\ }\textbf {\bibinfo {volume} {82}},\ \bibinfo {pages} {3045--3067} (\bibinfo {year} {2010})}\BibitemShut {NoStop}%
\bibitem [{\citenamefont {Resta}(1994)}]{resta1994modern}%
  \BibitemOpen
  \bibfield  {author} {\bibinfo {author} {\bibfnamefont {Raffaele}\ \bibnamefont {Resta}},\ }\bibfield  {title} {\enquote {\bibinfo {title} {Modern theory of polarization in ferroelectrics},}\ }\href@noop {} {\bibfield  {journal} {\bibinfo  {journal} {Ferroelectrics}\ }\textbf {\bibinfo {volume} {151}},\ \bibinfo {pages} {49--58} (\bibinfo {year} {1994})}\BibitemShut {NoStop}%
\bibitem [{\citenamefont {Rabe}\ \emph {et~al.}(2007)\citenamefont {Rabe}, \citenamefont {Ahn},\ and\ \citenamefont {Triscone}}]{rabe2007physics}%
  \BibitemOpen
  \bibfield  {author} {\bibinfo {author} {\bibfnamefont {Karin~M}\ \bibnamefont {Rabe}}, \bibinfo {author} {\bibfnamefont {Charles~H}\ \bibnamefont {Ahn}}, \ and\ \bibinfo {author} {\bibfnamefont {Jean-Marc}\ \bibnamefont {Triscone}},\ }\href@noop {} {\emph {\bibinfo {title} {Physics of ferroelectrics: a modern perspective}}},\ Vol.\ \bibinfo {volume} {105}\ (\bibinfo  {publisher} {Springer Science \& Business Media},\ \bibinfo {year} {2007})\BibitemShut {NoStop}%
\bibitem [{\citenamefont {Qi}\ \emph {et~al.}(2008)\citenamefont {Qi}, \citenamefont {Hughes},\ and\ \citenamefont {Zhang}}]{qi2008topological}%
  \BibitemOpen
  \bibfield  {author} {\bibinfo {author} {\bibfnamefont {Xiao-Liang}\ \bibnamefont {Qi}}, \bibinfo {author} {\bibfnamefont {Taylor~L}\ \bibnamefont {Hughes}}, \ and\ \bibinfo {author} {\bibfnamefont {Shou-Cheng}\ \bibnamefont {Zhang}},\ }\bibfield  {title} {\enquote {\bibinfo {title} {Topological field theory of time-reversal invariant insulators},}\ }\href@noop {} {\bibfield  {journal} {\bibinfo  {journal} {Physical Review B}\ }\textbf {\bibinfo {volume} {78}},\ \bibinfo {pages} {195424} (\bibinfo {year} {2008})}\BibitemShut {NoStop}%
\bibitem [{\citenamefont {Fu}\ and\ \citenamefont {Kane}(2007)}]{fu2007topological}%
  \BibitemOpen
  \bibfield  {author} {\bibinfo {author} {\bibfnamefont {Liang}\ \bibnamefont {Fu}}\ and\ \bibinfo {author} {\bibfnamefont {Charles~L}\ \bibnamefont {Kane}},\ }\bibfield  {title} {\enquote {\bibinfo {title} {Topological insulators with inversion symmetry},}\ }\href@noop {} {\bibfield  {journal} {\bibinfo  {journal} {Physical Review B}\ }\textbf {\bibinfo {volume} {76}},\ \bibinfo {pages} {045302} (\bibinfo {year} {2007})}\BibitemShut {NoStop}%
\bibitem [{\citenamefont {Schindler}\ \emph {et~al.}(2018)\citenamefont {Schindler}, \citenamefont {Cook}, \citenamefont {Vergniory}, \citenamefont {Wang}, \citenamefont {Parkin}, \citenamefont {Bernevig},\ and\ \citenamefont {Neupert}}]{schindler2018higher}%
  \BibitemOpen
  \bibfield  {author} {\bibinfo {author} {\bibfnamefont {Frank}\ \bibnamefont {Schindler}}, \bibinfo {author} {\bibfnamefont {Ashley~M}\ \bibnamefont {Cook}}, \bibinfo {author} {\bibfnamefont {Maia~G}\ \bibnamefont {Vergniory}}, \bibinfo {author} {\bibfnamefont {Zhijun}\ \bibnamefont {Wang}}, \bibinfo {author} {\bibfnamefont {Stuart~SP}\ \bibnamefont {Parkin}}, \bibinfo {author} {\bibfnamefont {B~Andrei}\ \bibnamefont {Bernevig}}, \ and\ \bibinfo {author} {\bibfnamefont {Titus}\ \bibnamefont {Neupert}},\ }\bibfield  {title} {\enquote {\bibinfo {title} {Higher-order topological insulators},}\ }\href@noop {} {\bibfield  {journal} {\bibinfo  {journal} {Science advances}\ }\textbf {\bibinfo {volume} {4}},\ \bibinfo {pages} {eaat0346} (\bibinfo {year} {2018})}\BibitemShut {NoStop}%
\bibitem [{\citenamefont {Khalaf}\ \emph {et~al.}(2018)\citenamefont {Khalaf}, \citenamefont {Po}, \citenamefont {Vishwanath},\ and\ \citenamefont {Watanabe}}]{khalaf2018symmetry}%
  \BibitemOpen
  \bibfield  {author} {\bibinfo {author} {\bibfnamefont {Eslam}\ \bibnamefont {Khalaf}}, \bibinfo {author} {\bibfnamefont {Hoi~Chun}\ \bibnamefont {Po}}, \bibinfo {author} {\bibfnamefont {Ashvin}\ \bibnamefont {Vishwanath}}, \ and\ \bibinfo {author} {\bibfnamefont {Haruki}\ \bibnamefont {Watanabe}},\ }\bibfield  {title} {\enquote {\bibinfo {title} {Symmetry indicators and anomalous surface states of topological crystalline insulators},}\ }\href@noop {} {\bibfield  {journal} {\bibinfo  {journal} {Physical Review X}\ }\textbf {\bibinfo {volume} {8}},\ \bibinfo {pages} {031070} (\bibinfo {year} {2018})}\BibitemShut {NoStop}%
\bibitem [{\citenamefont {Young}\ \emph {et~al.}(2012)\citenamefont {Young}, \citenamefont {Zaheer}, \citenamefont {Teo}, \citenamefont {Kane}, \citenamefont {Mele},\ and\ \citenamefont {Rappe}}]{young2012dirac}%
  \BibitemOpen
  \bibfield  {author} {\bibinfo {author} {\bibfnamefont {Steve~M}\ \bibnamefont {Young}}, \bibinfo {author} {\bibfnamefont {Saad}\ \bibnamefont {Zaheer}}, \bibinfo {author} {\bibfnamefont {Jeffrey~CY}\ \bibnamefont {Teo}}, \bibinfo {author} {\bibfnamefont {Charles~L}\ \bibnamefont {Kane}}, \bibinfo {author} {\bibfnamefont {Eugene~J}\ \bibnamefont {Mele}}, \ and\ \bibinfo {author} {\bibfnamefont {Andrew~M}\ \bibnamefont {Rappe}},\ }\bibfield  {title} {\enquote {\bibinfo {title} {Dirac semimetal in three dimensions},}\ }\href@noop {} {\bibfield  {journal} {\bibinfo  {journal} {Physical Review Letters}\ }\textbf {\bibinfo {volume} {108}},\ \bibinfo {pages} {140405} (\bibinfo {year} {2012})}\BibitemShut {NoStop}%
\bibitem [{\citenamefont {Yang}\ and\ \citenamefont {Nagaosa}(2014)}]{yang2014classification}%
  \BibitemOpen
  \bibfield  {author} {\bibinfo {author} {\bibfnamefont {Bohm-Jung}\ \bibnamefont {Yang}}\ and\ \bibinfo {author} {\bibfnamefont {Naoto}\ \bibnamefont {Nagaosa}},\ }\bibfield  {title} {\enquote {\bibinfo {title} {{Classification of stable three-dimensional Dirac semimetals with nontrivial topology}},}\ }\href@noop {} {\bibfield  {journal} {\bibinfo  {journal} {Nature Communications}\ }\textbf {\bibinfo {volume} {5}},\ \bibinfo {pages} {4898} (\bibinfo {year} {2014})}\BibitemShut {NoStop}%
\bibitem [{\citenamefont {Wieder}\ \emph {et~al.}(2020)\citenamefont {Wieder}, \citenamefont {Wang}, \citenamefont {Cano}, \citenamefont {Dai}, \citenamefont {Schoop}, \citenamefont {Bradlyn},\ and\ \citenamefont {Bernevig}}]{wieder2020strong}%
  \BibitemOpen
  \bibfield  {author} {\bibinfo {author} {\bibfnamefont {Benjamin~J}\ \bibnamefont {Wieder}}, \bibinfo {author} {\bibfnamefont {Zhijun}\ \bibnamefont {Wang}}, \bibinfo {author} {\bibfnamefont {Jennifer}\ \bibnamefont {Cano}}, \bibinfo {author} {\bibfnamefont {Xi}~\bibnamefont {Dai}}, \bibinfo {author} {\bibfnamefont {Leslie~M}\ \bibnamefont {Schoop}}, \bibinfo {author} {\bibfnamefont {Barry}\ \bibnamefont {Bradlyn}}, \ and\ \bibinfo {author} {\bibfnamefont {B~Andrei}\ \bibnamefont {Bernevig}},\ }\bibfield  {title} {\enquote {\bibinfo {title} {{Strong and fragile topological Dirac semimetals with higher-order Fermi arcs}},}\ }\href@noop {} {\bibfield  {journal} {\bibinfo  {journal} {Nature Communications}\ }\textbf {\bibinfo {volume} {11}},\ \bibinfo {pages} {627} (\bibinfo {year} {2020})}\BibitemShut {NoStop}%
\bibitem [{\citenamefont {Wang}\ \emph {et~al.}(2021)\citenamefont {Wang}, \citenamefont {Wu},\ and\ \citenamefont {An}}]{wang2021engineering}%
  \BibitemOpen
  \bibfield  {author} {\bibinfo {author} {\bibfnamefont {Bao-Qin}\ \bibnamefont {Wang}}, \bibinfo {author} {\bibfnamefont {Hong}\ \bibnamefont {Wu}}, \ and\ \bibinfo {author} {\bibfnamefont {Jun-Hong}\ \bibnamefont {An}},\ }\bibfield  {title} {\enquote {\bibinfo {title} {Engineering exotic second-order topological semimetals by periodic driving},}\ }\href@noop {} {\bibfield  {journal} {\bibinfo  {journal} {Physical Review B}\ }\textbf {\bibinfo {volume} {104}},\ \bibinfo {pages} {205117} (\bibinfo {year} {2021})}\BibitemShut {NoStop}%
\bibitem [{\citenamefont {Chen}\ \emph {et~al.}(2023)\citenamefont {Chen}, \citenamefont {Zhou},\ and\ \citenamefont {Xu}}]{chen2023quasicrystalline}%
  \BibitemOpen
  \bibfield  {author} {\bibinfo {author} {\bibfnamefont {Rui}\ \bibnamefont {Chen}}, \bibinfo {author} {\bibfnamefont {Bin}\ \bibnamefont {Zhou}}, \ and\ \bibinfo {author} {\bibfnamefont {Dong-Hui}\ \bibnamefont {Xu}},\ }\bibfield  {title} {\enquote {\bibinfo {title} {Quasicrystalline second-order topological semimetals},}\ }\href@noop {} {\bibfield  {journal} {\bibinfo  {journal} {Physical Review B}\ }\textbf {\bibinfo {volume} {108}},\ \bibinfo {pages} {195306} (\bibinfo {year} {2023})}\BibitemShut {NoStop}%
\bibitem [{\citenamefont {Wang}\ \emph {et~al.}(2023)\citenamefont {Wang}, \citenamefont {Wang}, \citenamefont {Hu}, \citenamefont {Bongiovanni}, \citenamefont {Juki{\'c}}, \citenamefont {Tang}, \citenamefont {Song}, \citenamefont {Morandotti}, \citenamefont {Chen},\ and\ \citenamefont {Buljan}}]{wang2023sub}%
  \BibitemOpen
  \bibfield  {author} {\bibinfo {author} {\bibfnamefont {Ziteng}\ \bibnamefont {Wang}}, \bibinfo {author} {\bibfnamefont {Xiangdong}\ \bibnamefont {Wang}}, \bibinfo {author} {\bibfnamefont {Zhichan}\ \bibnamefont {Hu}}, \bibinfo {author} {\bibfnamefont {Domenico}\ \bibnamefont {Bongiovanni}}, \bibinfo {author} {\bibfnamefont {Dario}\ \bibnamefont {Juki{\'c}}}, \bibinfo {author} {\bibfnamefont {Liqin}\ \bibnamefont {Tang}}, \bibinfo {author} {\bibfnamefont {Daohong}\ \bibnamefont {Song}}, \bibinfo {author} {\bibfnamefont {Roberto}\ \bibnamefont {Morandotti}}, \bibinfo {author} {\bibfnamefont {Zhigang}\ \bibnamefont {Chen}}, \ and\ \bibinfo {author} {\bibfnamefont {Hrvoje}\ \bibnamefont {Buljan}},\ }\bibfield  {title} {\enquote {\bibinfo {title} {Sub-symmetry-protected topological states},}\ }\href@noop {} {\bibfield  {journal} {\bibinfo  {journal} {Nature Physics}\ }\textbf {\bibinfo {volume} {19}},\ \bibinfo {pages} {992--998} (\bibinfo {year} {2023})}\BibitemShut {NoStop}%
\bibitem [{\citenamefont {Kang}\ \emph {et~al.}(2024)\citenamefont {Kang}, \citenamefont {Lee},\ and\ \citenamefont {Cheon}}]{kang2024subsymmetry}%
  \BibitemOpen
  \bibfield  {author} {\bibinfo {author} {\bibfnamefont {Myungjun}\ \bibnamefont {Kang}}, \bibinfo {author} {\bibfnamefont {Mingyu}\ \bibnamefont {Lee}}, \ and\ \bibinfo {author} {\bibfnamefont {Sangmo}\ \bibnamefont {Cheon}},\ }\bibfield  {title} {\enquote {\bibinfo {title} {Subsymmetry protected topology in topological insulators and superconductors},}\ }\href@noop {} {\bibfield  {journal} {\bibinfo  {journal} {Physical Review Research}\ }\textbf {\bibinfo {volume} {6}},\ \bibinfo {pages} {033323} (\bibinfo {year} {2024})}\BibitemShut {NoStop}%
\bibitem [{\citenamefont {Verma}\ and\ \citenamefont {Park}(2024)}]{verma2024non}%
  \BibitemOpen
  \bibfield  {author} {\bibinfo {author} {\bibfnamefont {Sonu}\ \bibnamefont {Verma}}\ and\ \bibinfo {author} {\bibfnamefont {Moon~Jip}\ \bibnamefont {Park}},\ }\bibfield  {title} {\enquote {\bibinfo {title} {Non-bloch band theory of subsymmetry-protected topological phases},}\ }\href@noop {} {\bibfield  {journal} {\bibinfo  {journal} {Physical Review B}\ }\textbf {\bibinfo {volume} {110}},\ \bibinfo {pages} {035424} (\bibinfo {year} {2024})}\BibitemShut {NoStop}%
\bibitem [{\citenamefont {Benalcazar}\ \emph {et~al.}(2017)\citenamefont {Benalcazar}, \citenamefont {Bernevig},\ and\ \citenamefont {Hughes}}]{benalcazar2017quantized}%
  \BibitemOpen
  \bibfield  {author} {\bibinfo {author} {\bibfnamefont {Wladimir~A}\ \bibnamefont {Benalcazar}}, \bibinfo {author} {\bibfnamefont {B~Andrei}\ \bibnamefont {Bernevig}}, \ and\ \bibinfo {author} {\bibfnamefont {Taylor~L}\ \bibnamefont {Hughes}},\ }\bibfield  {title} {\enquote {\bibinfo {title} {Quantized electric multipole insulators},}\ }\href@noop {} {\bibfield  {journal} {\bibinfo  {journal} {Science}\ }\textbf {\bibinfo {volume} {357}},\ \bibinfo {pages} {61--66} (\bibinfo {year} {2017})}\BibitemShut {NoStop}%
\bibitem [{\citenamefont {Drost}\ \emph {et~al.}(2017)\citenamefont {Drost}, \citenamefont {Ojanen}, \citenamefont {Harju},\ and\ \citenamefont {Liljeroth}}]{drost2017topological}%
  \BibitemOpen
  \bibfield  {author} {\bibinfo {author} {\bibfnamefont {Robert}\ \bibnamefont {Drost}}, \bibinfo {author} {\bibfnamefont {Teemu}\ \bibnamefont {Ojanen}}, \bibinfo {author} {\bibfnamefont {Ari}\ \bibnamefont {Harju}}, \ and\ \bibinfo {author} {\bibfnamefont {Peter}\ \bibnamefont {Liljeroth}},\ }\bibfield  {title} {\enquote {\bibinfo {title} {Topological states in engineered atomic lattices},}\ }\href@noop {} {\bibfield  {journal} {\bibinfo  {journal} {Nature Physics}\ }\textbf {\bibinfo {volume} {13}},\ \bibinfo {pages} {668--671} (\bibinfo {year} {2017})}\BibitemShut {NoStop}%
\bibitem [{\citenamefont {Huda}\ \emph {et~al.}(2020)\citenamefont {Huda}, \citenamefont {Kezilebieke}, \citenamefont {Ojanen}, \citenamefont {Drost},\ and\ \citenamefont {Liljeroth}}]{huda2020tuneable}%
  \BibitemOpen
  \bibfield  {author} {\bibinfo {author} {\bibfnamefont {Md~Nurul}\ \bibnamefont {Huda}}, \bibinfo {author} {\bibfnamefont {Shawulienu}\ \bibnamefont {Kezilebieke}}, \bibinfo {author} {\bibfnamefont {Teemu}\ \bibnamefont {Ojanen}}, \bibinfo {author} {\bibfnamefont {Robert}\ \bibnamefont {Drost}}, \ and\ \bibinfo {author} {\bibfnamefont {Peter}\ \bibnamefont {Liljeroth}},\ }\bibfield  {title} {\enquote {\bibinfo {title} {Tuneable topological domain wall states in engineered atomic chains},}\ }\href@noop {} {\bibfield  {journal} {\bibinfo  {journal} {Npj quantum materials}\ }\textbf {\bibinfo {volume} {5}},\ \bibinfo {pages} {17} (\bibinfo {year} {2020})}\BibitemShut {NoStop}%
\bibitem [{\citenamefont {Meier}\ \emph {et~al.}(2016)\citenamefont {Meier}, \citenamefont {An},\ and\ \citenamefont {Gadway}}]{meier2016observation}%
  \BibitemOpen
  \bibfield  {author} {\bibinfo {author} {\bibfnamefont {Eric~J}\ \bibnamefont {Meier}}, \bibinfo {author} {\bibfnamefont {Fangzhao~Alex}\ \bibnamefont {An}}, \ and\ \bibinfo {author} {\bibfnamefont {Bryce}\ \bibnamefont {Gadway}},\ }\bibfield  {title} {\enquote {\bibinfo {title} {{Observation of the topological soliton state in the Su-Schrieffer-Heeger model}},}\ }\href@noop {} {\bibfield  {journal} {\bibinfo  {journal} {Nature Communications}\ }\textbf {\bibinfo {volume} {7}},\ \bibinfo {pages} {13986} (\bibinfo {year} {2016})}\BibitemShut {NoStop}%
\bibitem [{\citenamefont {Ozawa}\ \emph {et~al.}(2019)\citenamefont {Ozawa}, \citenamefont {Price}, \citenamefont {Amo}, \citenamefont {Goldman}, \citenamefont {Hafezi}, \citenamefont {Lu}, \citenamefont {Rechtsman}, \citenamefont {Schuster}, \citenamefont {Simon}, \citenamefont {Zilberberg} \emph {et~al.}}]{ozawa2019topological}%
  \BibitemOpen
  \bibfield  {author} {\bibinfo {author} {\bibfnamefont {Tomoki}\ \bibnamefont {Ozawa}}, \bibinfo {author} {\bibfnamefont {Hannah~M}\ \bibnamefont {Price}}, \bibinfo {author} {\bibfnamefont {Alberto}\ \bibnamefont {Amo}}, \bibinfo {author} {\bibfnamefont {Nathan}\ \bibnamefont {Goldman}}, \bibinfo {author} {\bibfnamefont {Mohammad}\ \bibnamefont {Hafezi}}, \bibinfo {author} {\bibfnamefont {Ling}\ \bibnamefont {Lu}}, \bibinfo {author} {\bibfnamefont {Mikael~C}\ \bibnamefont {Rechtsman}}, \bibinfo {author} {\bibfnamefont {David}\ \bibnamefont {Schuster}}, \bibinfo {author} {\bibfnamefont {Jonathan}\ \bibnamefont {Simon}}, \bibinfo {author} {\bibfnamefont {Oded}\ \bibnamefont {Zilberberg}},  \emph {et~al.},\ }\bibfield  {title} {\enquote {\bibinfo {title} {Topological photonics},}\ }\href@noop {} {\bibfield  {journal} {\bibinfo  {journal} {Reviews of Modern Physics}\ }\textbf {\bibinfo {volume} {91}},\ \bibinfo {pages} {015006} (\bibinfo {year} {2019})}\BibitemShut {NoStop}%
\bibitem [{\citenamefont {Roy}\ \emph {et~al.}(2021)\citenamefont {Roy}, \citenamefont {Mishra}, \citenamefont {Tanatar},\ and\ \citenamefont {Basu}}]{roy2021reentrant}%
  \BibitemOpen
  \bibfield  {author} {\bibinfo {author} {\bibfnamefont {Shilpi}\ \bibnamefont {Roy}}, \bibinfo {author} {\bibfnamefont {Tapan}\ \bibnamefont {Mishra}}, \bibinfo {author} {\bibfnamefont {Bilal}\ \bibnamefont {Tanatar}}, \ and\ \bibinfo {author} {\bibfnamefont {Saurabh}\ \bibnamefont {Basu}},\ }\bibfield  {title} {\enquote {\bibinfo {title} {Reentrant localization transition in a quasiperiodic chain},}\ }\href@noop {} {\bibfield  {journal} {\bibinfo  {journal} {Physical Review Letters}\ }\textbf {\bibinfo {volume} {126}},\ \bibinfo {pages} {106803} (\bibinfo {year} {2021})}\BibitemShut {NoStop}%
\bibitem [{\citenamefont {Roy}\ \emph {et~al.}(2023)\citenamefont {Roy}, \citenamefont {Nabi},\ and\ \citenamefont {Basu}}]{roy2023critical}%
  \BibitemOpen
  \bibfield  {author} {\bibinfo {author} {\bibfnamefont {Shilpi}\ \bibnamefont {Roy}}, \bibinfo {author} {\bibfnamefont {Sk~Noor}\ \bibnamefont {Nabi}}, \ and\ \bibinfo {author} {\bibfnamefont {Saurabh}\ \bibnamefont {Basu}},\ }\bibfield  {title} {\enquote {\bibinfo {title} {Critical and topological phases of dimerized kitaev chain in presence of quasiperiodic potential},}\ }\href@noop {} {\bibfield  {journal} {\bibinfo  {journal} {Physical Review B}\ }\textbf {\bibinfo {volume} {107}},\ \bibinfo {pages} {014202} (\bibinfo {year} {2023})}\BibitemShut {NoStop}%
\bibitem [{\citenamefont {Schindler}(2020)}]{schindler2020dirac}%
  \BibitemOpen
  \bibfield  {author} {\bibinfo {author} {\bibfnamefont {Frank}\ \bibnamefont {Schindler}},\ }\bibfield  {title} {\enquote {\bibinfo {title} {Dirac equation perspective on higher-order topological insulators},}\ }\href@noop {} {\bibfield  {journal} {\bibinfo  {journal} {Journal of Applied Physics}\ }\textbf {\bibinfo {volume} {128}} (\bibinfo {year} {2020})}\BibitemShut {NoStop}%
\bibitem [{\citenamefont {Schnyder}\ \emph {et~al.}(2008)\citenamefont {Schnyder}, \citenamefont {Ryu}, \citenamefont {Furusaki},\ and\ \citenamefont {Ludwig}}]{schnyder2008classification}%
  \BibitemOpen
  \bibfield  {author} {\bibinfo {author} {\bibfnamefont {Andreas~P}\ \bibnamefont {Schnyder}}, \bibinfo {author} {\bibfnamefont {Shinsei}\ \bibnamefont {Ryu}}, \bibinfo {author} {\bibfnamefont {Akira}\ \bibnamefont {Furusaki}}, \ and\ \bibinfo {author} {\bibfnamefont {Andreas~WW}\ \bibnamefont {Ludwig}},\ }\bibfield  {title} {\enquote {\bibinfo {title} {Classification of topological insulators and superconductors in three spatial dimensions},}\ }\href@noop {} {\bibfield  {journal} {\bibinfo  {journal} {Physical Review B}\ }\textbf {\bibinfo {volume} {78}},\ \bibinfo {pages} {195125} (\bibinfo {year} {2008})}\BibitemShut {NoStop}%
\bibitem [{\citenamefont {Chiu}\ \emph {et~al.}(2016)\citenamefont {Chiu}, \citenamefont {Teo}, \citenamefont {Schnyder},\ and\ \citenamefont {Ryu}}]{chiu2016classification}%
  \BibitemOpen
  \bibfield  {author} {\bibinfo {author} {\bibfnamefont {Ching-Kai}\ \bibnamefont {Chiu}}, \bibinfo {author} {\bibfnamefont {Jeffrey~CY}\ \bibnamefont {Teo}}, \bibinfo {author} {\bibfnamefont {Andreas~P}\ \bibnamefont {Schnyder}}, \ and\ \bibinfo {author} {\bibfnamefont {Shinsei}\ \bibnamefont {Ryu}},\ }\bibfield  {title} {\enquote {\bibinfo {title} {Classification of topological quantum matter with symmetries},}\ }\href@noop {} {\bibfield  {journal} {\bibinfo  {journal} {Reviews of Modern Physics}\ }\textbf {\bibinfo {volume} {88}},\ \bibinfo {pages} {035005} (\bibinfo {year} {2016})}\BibitemShut {NoStop}%
\bibitem [{\citenamefont {Jackiw}\ and\ \citenamefont {Rebbi}(1976)}]{jackiw1976solitons}%
  \BibitemOpen
  \bibfield  {author} {\bibinfo {author} {\bibfnamefont {Roman}\ \bibnamefont {Jackiw}}\ and\ \bibinfo {author} {\bibfnamefont {Cl{\'a}udio}\ \bibnamefont {Rebbi}},\ }\bibfield  {title} {\enquote {\bibinfo {title} {Solitons with fermion number $\frac{1}{2}$},}\ }\href@noop {} {\bibfield  {journal} {\bibinfo  {journal} {Physical Review D}\ }\textbf {\bibinfo {volume} {13}},\ \bibinfo {pages} {3398} (\bibinfo {year} {1976})}\BibitemShut {NoStop}%
\bibitem [{\citenamefont {Jackiw}\ and\ \citenamefont {Schrieffer}(1981)}]{jackiw1981solitons}%
  \BibitemOpen
  \bibfield  {author} {\bibinfo {author} {\bibfnamefont {Roman}\ \bibnamefont {Jackiw}}\ and\ \bibinfo {author} {\bibfnamefont {John~Robert}\ \bibnamefont {Schrieffer}},\ }\bibfield  {title} {\enquote {\bibinfo {title} {Solitons with fermion number $\frac{1}{2}$ in condensed matter and relativistic field theories},}\ }\href@noop {} {\bibfield  {journal} {\bibinfo  {journal} {Nuclear Physics B}\ }\textbf {\bibinfo {volume} {190}},\ \bibinfo {pages} {253--265} (\bibinfo {year} {1981})}\BibitemShut {NoStop}%
\bibitem [{\citenamefont {Liu}\ \emph {et~al.}(2014)\citenamefont {Liu}, \citenamefont {Zhou}, \citenamefont {Zhang}, \citenamefont {Wang}, \citenamefont {Weng}, \citenamefont {Prabhakaran}, \citenamefont {Mo}, \citenamefont {Shen}, \citenamefont {Fang}, \citenamefont {Dai} \emph {et~al.}}]{liu2014discovery}%
  \BibitemOpen
  \bibfield  {author} {\bibinfo {author} {\bibfnamefont {ZK}~\bibnamefont {Liu}}, \bibinfo {author} {\bibfnamefont {Bo}~\bibnamefont {Zhou}}, \bibinfo {author} {\bibfnamefont {Yong}\ \bibnamefont {Zhang}}, \bibinfo {author} {\bibfnamefont {ZJ}~\bibnamefont {Wang}}, \bibinfo {author} {\bibfnamefont {HM}~\bibnamefont {Weng}}, \bibinfo {author} {\bibfnamefont {Dharmalingam}\ \bibnamefont {Prabhakaran}}, \bibinfo {author} {\bibfnamefont {S-K}\ \bibnamefont {Mo}}, \bibinfo {author} {\bibfnamefont {ZX}~\bibnamefont {Shen}}, \bibinfo {author} {\bibfnamefont {Zhong}\ \bibnamefont {Fang}}, \bibinfo {author} {\bibfnamefont {Xi}~\bibnamefont {Dai}},  \emph {et~al.},\ }\bibfield  {title} {\enquote {\bibinfo {title} {{Discovery of a three-dimensional topological Dirac semimetal, Na\textsubscript{3}Bi}},}\ }\href@noop {} {\bibfield  {journal} {\bibinfo  {journal} {Science}\ }\textbf {\bibinfo {volume} {343}},\ \bibinfo {pages} {864--867} (\bibinfo {year} {2014})}\BibitemShut {NoStop}%
\bibitem [{\citenamefont {Borisenko}\ \emph {et~al.}(2014)\citenamefont {Borisenko}, \citenamefont {Gibson}, \citenamefont {Evtushinsky}, \citenamefont {Zabolotnyy}, \citenamefont {B{\"u}chner},\ and\ \citenamefont {Cava}}]{borisenko2014experimental}%
  \BibitemOpen
  \bibfield  {author} {\bibinfo {author} {\bibfnamefont {Sergey}\ \bibnamefont {Borisenko}}, \bibinfo {author} {\bibfnamefont {Quinn}\ \bibnamefont {Gibson}}, \bibinfo {author} {\bibfnamefont {Danil}\ \bibnamefont {Evtushinsky}}, \bibinfo {author} {\bibfnamefont {Volodymyr}\ \bibnamefont {Zabolotnyy}}, \bibinfo {author} {\bibfnamefont {Bernd}\ \bibnamefont {B{\"u}chner}}, \ and\ \bibinfo {author} {\bibfnamefont {Robert~J}\ \bibnamefont {Cava}},\ }\bibfield  {title} {\enquote {\bibinfo {title} {{Experimental realization of a three-dimensional Dirac semimetal}},}\ }\href@noop {} {\bibfield  {journal} {\bibinfo  {journal} {Physical Review Letters}\ }\textbf {\bibinfo {volume} {113}},\ \bibinfo {pages} {027603} (\bibinfo {year} {2014})}\BibitemShut {NoStop}%
\bibitem [{\citenamefont {Neupane}\ \emph {et~al.}(2014)\citenamefont {Neupane}, \citenamefont {Xu}, \citenamefont {Sankar}, \citenamefont {Alidoust}, \citenamefont {Bian}, \citenamefont {Liu}, \citenamefont {Belopolski}, \citenamefont {Chang}, \citenamefont {Jeng}, \citenamefont {Lin} \emph {et~al.}}]{neupane2014observation}%
  \BibitemOpen
  \bibfield  {author} {\bibinfo {author} {\bibfnamefont {Madhab}\ \bibnamefont {Neupane}}, \bibinfo {author} {\bibfnamefont {Su-Yang}\ \bibnamefont {Xu}}, \bibinfo {author} {\bibfnamefont {Raman}\ \bibnamefont {Sankar}}, \bibinfo {author} {\bibfnamefont {Nasser}\ \bibnamefont {Alidoust}}, \bibinfo {author} {\bibfnamefont {Guang}\ \bibnamefont {Bian}}, \bibinfo {author} {\bibfnamefont {Chang}\ \bibnamefont {Liu}}, \bibinfo {author} {\bibfnamefont {Ilya}\ \bibnamefont {Belopolski}}, \bibinfo {author} {\bibfnamefont {Tay-Rong}\ \bibnamefont {Chang}}, \bibinfo {author} {\bibfnamefont {Horng-Tay}\ \bibnamefont {Jeng}}, \bibinfo {author} {\bibfnamefont {Hsin}\ \bibnamefont {Lin}},  \emph {et~al.},\ }\bibfield  {title} {\enquote {\bibinfo {title} {{Observation of a three-dimensional topological Dirac semimetal phase in high-mobility Cd\textsubscript{3}As\textsubscript{2}}},}\ }\href@noop {} {\bibfield  {journal} {\bibinfo  {journal} {Nature Communications}\ }\textbf {\bibinfo {volume} {5}},\ \bibinfo {pages} {3786}
  (\bibinfo {year} {2014})}\BibitemShut {NoStop}%
\bibitem [{\citenamefont {Ganeshan}\ and\ \citenamefont {Das~Sarma}(2015)}]{ganeshan2015constructing}%
  \BibitemOpen
  \bibfield  {author} {\bibinfo {author} {\bibfnamefont {Sriram}\ \bibnamefont {Ganeshan}}\ and\ \bibinfo {author} {\bibfnamefont {S}~\bibnamefont {Das~Sarma}},\ }\bibfield  {title} {\enquote {\bibinfo {title} {{Constructing a Weyl semimetal by stacking one-dimensional topological phases}},}\ }\href@noop {} {\bibfield  {journal} {\bibinfo  {journal} {Physical Review B}\ }\textbf {\bibinfo {volume} {91}},\ \bibinfo {pages} {125438} (\bibinfo {year} {2015})}\BibitemShut {NoStop}%
\bibitem [{\citenamefont {Ezawa}(2018)}]{ezawa2018higher}%
  \BibitemOpen
  \bibfield  {author} {\bibinfo {author} {\bibfnamefont {Motohiko}\ \bibnamefont {Ezawa}},\ }\bibfield  {title} {\enquote {\bibinfo {title} {Higher-order topological electric circuits and topological corner resonance on the breathing kagome and pyrochlore lattices},}\ }\href@noop {} {\bibfield  {journal} {\bibinfo  {journal} {Physical Review B}\ }\textbf {\bibinfo {volume} {98}},\ \bibinfo {pages} {201402} (\bibinfo {year} {2018})}\BibitemShut {NoStop}%
\bibitem [{\citenamefont {Imhof}\ \emph {et~al.}(2018)\citenamefont {Imhof}, \citenamefont {Berger}, \citenamefont {Bayer}, \citenamefont {Brehm}, \citenamefont {Molenkamp}, \citenamefont {Kiessling}, \citenamefont {Schindler}, \citenamefont {Lee}, \citenamefont {Greiter}, \citenamefont {Neupert} \emph {et~al.}}]{imhof2018topolectrical}%
  \BibitemOpen
  \bibfield  {author} {\bibinfo {author} {\bibfnamefont {Stefan}\ \bibnamefont {Imhof}}, \bibinfo {author} {\bibfnamefont {Christian}\ \bibnamefont {Berger}}, \bibinfo {author} {\bibfnamefont {Florian}\ \bibnamefont {Bayer}}, \bibinfo {author} {\bibfnamefont {Johannes}\ \bibnamefont {Brehm}}, \bibinfo {author} {\bibfnamefont {Laurens~W}\ \bibnamefont {Molenkamp}}, \bibinfo {author} {\bibfnamefont {Tobias}\ \bibnamefont {Kiessling}}, \bibinfo {author} {\bibfnamefont {Frank}\ \bibnamefont {Schindler}}, \bibinfo {author} {\bibfnamefont {Ching~Hua}\ \bibnamefont {Lee}}, \bibinfo {author} {\bibfnamefont {Martin}\ \bibnamefont {Greiter}}, \bibinfo {author} {\bibfnamefont {Titus}\ \bibnamefont {Neupert}},  \emph {et~al.},\ }\bibfield  {title} {\enquote {\bibinfo {title} {Topolectrical-circuit realization of topological corner modes},}\ }\href@noop {} {\bibfield  {journal} {\bibinfo  {journal} {Nature Physics}\ }\textbf {\bibinfo {volume} {14}},\ \bibinfo {pages} {925--929} (\bibinfo {year} {2018})}\BibitemShut
  {NoStop}%
\bibitem [{\citenamefont {Lee}\ \emph {et~al.}(2018)\citenamefont {Lee}, \citenamefont {Imhof}, \citenamefont {Berger}, \citenamefont {Bayer}, \citenamefont {Brehm}, \citenamefont {Molenkamp}, \citenamefont {Kiessling},\ and\ \citenamefont {Thomale}}]{lee2018topolectrical}%
  \BibitemOpen
  \bibfield  {author} {\bibinfo {author} {\bibfnamefont {Ching~Hua}\ \bibnamefont {Lee}}, \bibinfo {author} {\bibfnamefont {Stefan}\ \bibnamefont {Imhof}}, \bibinfo {author} {\bibfnamefont {Christian}\ \bibnamefont {Berger}}, \bibinfo {author} {\bibfnamefont {Florian}\ \bibnamefont {Bayer}}, \bibinfo {author} {\bibfnamefont {Johannes}\ \bibnamefont {Brehm}}, \bibinfo {author} {\bibfnamefont {Laurens~W}\ \bibnamefont {Molenkamp}}, \bibinfo {author} {\bibfnamefont {Tobias}\ \bibnamefont {Kiessling}}, \ and\ \bibinfo {author} {\bibfnamefont {Ronny}\ \bibnamefont {Thomale}},\ }\bibfield  {title} {\enquote {\bibinfo {title} {Topolectrical circuits},}\ }\href@noop {} {\bibfield  {journal} {\bibinfo  {journal} {Communications Physics}\ }\textbf {\bibinfo {volume} {1}},\ \bibinfo {pages} {39} (\bibinfo {year} {2018})}\BibitemShut {NoStop}%
\bibitem [{\citenamefont {Dong}\ \emph {et~al.}(2021)\citenamefont {Dong}, \citenamefont {Juri{\v{c}}i{\'c}},\ and\ \citenamefont {Roy}}]{dong2021topolectric}%
  \BibitemOpen
  \bibfield  {author} {\bibinfo {author} {\bibfnamefont {Junkai}\ \bibnamefont {Dong}}, \bibinfo {author} {\bibfnamefont {Vladimir}\ \bibnamefont {Juri{\v{c}}i{\'c}}}, \ and\ \bibinfo {author} {\bibfnamefont {Bitan}\ \bibnamefont {Roy}},\ }\bibfield  {title} {\enquote {\bibinfo {title} {Topolectric circuits: Theory and construction},}\ }\href@noop {} {\bibfield  {journal} {\bibinfo  {journal} {Physical Review Research}\ }\textbf {\bibinfo {volume} {3}},\ \bibinfo {pages} {023056} (\bibinfo {year} {2021})}\BibitemShut {NoStop}%
\bibitem [{\citenamefont {Xie}\ \emph {et~al.}(2018)\citenamefont {Xie}, \citenamefont {Wang}, \citenamefont {Wang}, \citenamefont {Zhu}, \citenamefont {Jiang}, \citenamefont {Lu},\ and\ \citenamefont {Chen}}]{xie2018second}%
  \BibitemOpen
  \bibfield  {author} {\bibinfo {author} {\bibfnamefont {Bi-Ye}\ \bibnamefont {Xie}}, \bibinfo {author} {\bibfnamefont {Hong-Fei}\ \bibnamefont {Wang}}, \bibinfo {author} {\bibfnamefont {Hai-Xiao}\ \bibnamefont {Wang}}, \bibinfo {author} {\bibfnamefont {Xue-Yi}\ \bibnamefont {Zhu}}, \bibinfo {author} {\bibfnamefont {Jian-Hua}\ \bibnamefont {Jiang}}, \bibinfo {author} {\bibfnamefont {Ming-Hui}\ \bibnamefont {Lu}}, \ and\ \bibinfo {author} {\bibfnamefont {Yan-Feng}\ \bibnamefont {Chen}},\ }\bibfield  {title} {\enquote {\bibinfo {title} {Second-order photonic topological insulator with corner states},}\ }\href@noop {} {\bibfield  {journal} {\bibinfo  {journal} {Physical Review B}\ }\textbf {\bibinfo {volume} {98}},\ \bibinfo {pages} {205147} (\bibinfo {year} {2018})}\BibitemShut {NoStop}%
\bibitem [{\citenamefont {Li}\ \emph {et~al.}(2022)\citenamefont {Li}, \citenamefont {Li}, \citenamefont {Yan}, \citenamefont {Ye}, \citenamefont {Hu}, \citenamefont {Gong},\ and\ \citenamefont {Li}}]{li2022higher}%
  \BibitemOpen
  \bibfield  {author} {\bibinfo {author} {\bibfnamefont {Chu}\ \bibnamefont {Li}}, \bibinfo {author} {\bibfnamefont {Meng}\ \bibnamefont {Li}}, \bibinfo {author} {\bibfnamefont {Linyu}\ \bibnamefont {Yan}}, \bibinfo {author} {\bibfnamefont {Sheng}\ \bibnamefont {Ye}}, \bibinfo {author} {\bibfnamefont {Xiaoyong}\ \bibnamefont {Hu}}, \bibinfo {author} {\bibfnamefont {Qihuang}\ \bibnamefont {Gong}}, \ and\ \bibinfo {author} {\bibfnamefont {Yan}\ \bibnamefont {Li}},\ }\bibfield  {title} {\enquote {\bibinfo {title} {Higher-order topological biphoton corner states in two-dimensional photonic lattices},}\ }\href@noop {} {\bibfield  {journal} {\bibinfo  {journal} {Physical Review Research}\ }\textbf {\bibinfo {volume} {4}},\ \bibinfo {pages} {023049} (\bibinfo {year} {2022})}\BibitemShut {NoStop}%
\bibitem [{\citenamefont {Schulz}\ \emph {et~al.}(2022)\citenamefont {Schulz}, \citenamefont {Noh}, \citenamefont {Benalcazar}, \citenamefont {Bahl},\ and\ \citenamefont {von Freymann}}]{schulz2022photonic}%
  \BibitemOpen
  \bibfield  {author} {\bibinfo {author} {\bibfnamefont {Julian}\ \bibnamefont {Schulz}}, \bibinfo {author} {\bibfnamefont {Jiho}\ \bibnamefont {Noh}}, \bibinfo {author} {\bibfnamefont {Wladimir~A}\ \bibnamefont {Benalcazar}}, \bibinfo {author} {\bibfnamefont {Gaurav}\ \bibnamefont {Bahl}}, \ and\ \bibinfo {author} {\bibfnamefont {Georg}\ \bibnamefont {von Freymann}},\ }\bibfield  {title} {\enquote {\bibinfo {title} {Photonic quadrupole topological insulator using orbital-induced synthetic flux},}\ }\href@noop {} {\bibfield  {journal} {\bibinfo  {journal} {Nature Communications}\ }\textbf {\bibinfo {volume} {13}},\ \bibinfo {pages} {6597} (\bibinfo {year} {2022})}\BibitemShut {NoStop}%
\bibitem [{\citenamefont {El~Hassan}\ \emph {et~al.}(2019)\citenamefont {El~Hassan}, \citenamefont {Kunst}, \citenamefont {Moritz}, \citenamefont {Andler}, \citenamefont {Bergholtz},\ and\ \citenamefont {Bourennane}}]{el2019corner}%
  \BibitemOpen
  \bibfield  {author} {\bibinfo {author} {\bibfnamefont {Ashraf}\ \bibnamefont {El~Hassan}}, \bibinfo {author} {\bibfnamefont {Flore~K}\ \bibnamefont {Kunst}}, \bibinfo {author} {\bibfnamefont {Alexander}\ \bibnamefont {Moritz}}, \bibinfo {author} {\bibfnamefont {Guillermo}\ \bibnamefont {Andler}}, \bibinfo {author} {\bibfnamefont {Emil~J}\ \bibnamefont {Bergholtz}}, \ and\ \bibinfo {author} {\bibfnamefont {Mohamed}\ \bibnamefont {Bourennane}},\ }\bibfield  {title} {\enquote {\bibinfo {title} {Corner states of light in photonic waveguides},}\ }\href@noop {} {\bibfield  {journal} {\bibinfo  {journal} {Nature Photonics}\ }\textbf {\bibinfo {volume} {13}},\ \bibinfo {pages} {697--700} (\bibinfo {year} {2019})}\BibitemShut {NoStop}%
\bibitem [{\citenamefont {Zhou}\ \emph {et~al.}(2024)\citenamefont {Zhou}, \citenamefont {Liu}, \citenamefont {Wang}, \citenamefont {Li}, \citenamefont {Xie}, \citenamefont {Zhang}, \citenamefont {Mandal}, \citenamefont {Xi}, \citenamefont {Gao}, \citenamefont {Deng} \emph {et~al.}}]{zhou2024realization}%
  \BibitemOpen
  \bibfield  {author} {\bibinfo {author} {\bibfnamefont {Peiheng}\ \bibnamefont {Zhou}}, \bibinfo {author} {\bibfnamefont {Gui-Geng}\ \bibnamefont {Liu}}, \bibinfo {author} {\bibfnamefont {Zihao}\ \bibnamefont {Wang}}, \bibinfo {author} {\bibfnamefont {Shuwei}\ \bibnamefont {Li}}, \bibinfo {author} {\bibfnamefont {Qindong}\ \bibnamefont {Xie}}, \bibinfo {author} {\bibfnamefont {Yunpeng}\ \bibnamefont {Zhang}}, \bibinfo {author} {\bibfnamefont {Subhaskar}\ \bibnamefont {Mandal}}, \bibinfo {author} {\bibfnamefont {Xiang}\ \bibnamefont {Xi}}, \bibinfo {author} {\bibfnamefont {Zhen}\ \bibnamefont {Gao}}, \bibinfo {author} {\bibfnamefont {Longjiang}\ \bibnamefont {Deng}},  \emph {et~al.},\ }\bibfield  {title} {\enquote {\bibinfo {title} {Realization of a quadrupole topological insulator phase in a gyromagnetic photonic crystal},}\ }\href@noop {} {\bibfield  {journal} {\bibinfo  {journal} {National Science Review}\ }\textbf {\bibinfo {volume} {11}},\ \bibinfo {pages} {nwae121} (\bibinfo {year} {2024})}\BibitemShut
  {NoStop}%
\end{thebibliography}
\end{document}